\documentclass[preprint,double-spaced,floatfix]{revtex4}
\usepackage[utf8]{inputenc}
\usepackage{bbm}
\usepackage{mathrsfs}
\usepackage{epsfig,psfrag}
\usepackage[final]{pdfpages}
\usepackage{amsmath,amsfonts,amssymb}
\usepackage{graphics}
\usepackage{pstool}
\usepackage{anyfontsize}
\usepackage{rotating}
\usepackage[dvips, bookmarks, colorlinks=true, plainpages = false, citecolor = blue, linkcolor = blue, urlcolor = blue, filecolor = blue]{hyperref}
\newcommand{\ket}[1]{\ensuremath{\left\langle#1\right|}}
\newcommand{\bra}[1]{\ensuremath{\left|#1\right\rangle}}
\def\be{\begin{equation}}
\def\ee{\end{equation}}
\def\bea{\begin{eqnarray}}
\def\eea{\end{eqnarray}}

\newcommand{\Gcal}{\mathcal G}
\newcommand{\Fcal}{\mathcal F}

\begin{document}


\title{Thermally stable multipartite entanglements in the frustrated Heisenberg hexagon}

\author{Moumita Deb}\email{moumitadeb44@gmail.com}
\author{Asim Kumar Ghosh}
 \email{asimkumar96@yahoo.com}
\affiliation {Department of Physics, Jadavpur University, 
188 Raja Subodh Chandra Mallik Road, Kolkata 700032, India}

\begin{abstract}
Thermally stable quantum states with multipartite entanglements led by frustration 
are found in the antiferromagnetic spin-1/2 Heisenberg hexagon. 
The model has been solved exactly to obtain all analytic expressions of 
eigenvalues and eigenfunctions. 
Detection and characterizations 
for various types of entanglements have been carried out in terms of concurrence and 
entanglement witnesses based on several thermodynamic observables. 
Variations of entanglement properties with respect to 
temperature and frustration are discussed. 
Even though the frustration opposes the bipartite entanglement, it 
favors the multipartite entanglement. 
Entangled states exhibit robustness against the thermal effects 
in the presence of frustration 
and they are found to survive at any temperature. 
\end{abstract}

\maketitle

\section{INTRODUCTION}
The field of quantum information processing experiences a mammoth growth 
in the last two decades \cite{Amico,HorodeckisRMP,Guuhne}. 
Entanglement emerges as the most useful quantity among the 
quantum correlations through an extensive investigations in this field. 
Vast amounts of works involve in detection, 
characterization, distillation and quantification 
of entanglements associated in various quantum systems. 
Nowadays, there are plethora of ideas which pave the way to realize the  
more secure and faster information processing tools as well as the 
more stable and efficient quantum communication networks. 
These technological innovations include cryptography \cite{Ekert}, 
dense coding \cite{Bennett1}, teleportation \cite{Bennett2} and many more. 
Interacting spin models consist of both small clusters and long chains 
where the spins are interacting through the exchange 
interactions can serve as the platforms to verify the outcome 
of those proposals \cite{QSTNE}. 
Thermal stability of the entangled state, 
on the other hand,  
is the main concern to make those protocols operational at room temperature. 

Besides those achievements, quantum correlations exhibit dramatic 
changes in their values when the system undergoes a quantum phase 
transition (QPT) \cite{Sachdev,Somma,JVidal,Wu}.  
Again, value of those correlations can be obtained exactly for the spin 
models as well as the locations of QPTs can be identified more precisely. 
In addition, real materials are also available those could serve as the macroscopic 
realizations of any specific spin models. 
One of the example of such material is polyoxovanadate compound, 
(NHEt)$_3$ [V$_8^{\rm IV}$V$_4^{\rm V}$As$_8$O$_{40}$(H$_2$O)]$\cdot$H$_2$O \cite{Procissi}. 
The magnetic properties of this compound are faithfully explained by considering 
a four-spin cluster, in which four spin-1/2 degrees are arranged on  
the vertices of a square and they are interacting with the nearest one with 
isotropic antiferromagnetic (AFM) Heisenberg exchange couplings. 
QPT occurs at a definite point for this model in the presence of 
diagonal exchange interaction \cite{Bose}. 
Experimental evidence suggests that 
entanglement can affect the macroscopic properties of solids. 
The observed values of specific heat and magnetic susceptibility 
for the compound LiHO$_x$Y$_{1-x}$F$_4$ predict that those can be explained 
if entanglement of the relevant quantum states are considered explicitly \cite{Ghosh}. 
Thermodynamic observables of macroscopic system, like 
internal energy \cite{Dowling}, susceptibility \cite{Wiesniak} and 
structure factor \cite{Krammer} serve as the entanglement witness (EW), 
since the measurement of those quantities eventually leads to the detection of 
entanglement. For example, the magnetic structure of deuterated copper nitrate  
Cu(NO$_3$)$_2$2.5D$_2$O has been considered as composed of 
uncoupled spin-1/2 bond alternating AFM Heisenberg chains 
and for this material susceptibility acts as EW \cite{Brukner}. 
A rigorous study of entanglement properties for 
Heisenberg spin chains in the thermodynamic limit 
is a theoretical challenge since the eigenvalues and eigenfunctions are not known 
exactly in every case. In many cases, exact results are 
obtainable whenever the spin chain is mapped onto a spinless 
fermionic models \cite{Korepin,Mezzadri,Osborne,GVidal,Damski}. 
Bipartite and multipartite entanglement properties of various 
spin models have been investigated at finite temperatures 
\cite{Wooters_prl80,Wooters_pra63,Bruss,Glaser,Osterloh,Bose}. 
But the analytic derivations of all  
entanglement properties for an isolated cluster containing 
few spins is possible so far as it is exactly diagonalizable. 
Moreover, spin cluster with higher values of exchange 
strength can enhance the stability of entangled state at room temperature which is facing a 
real challenge nowadays. A spin-cluster material,  
copper carboxylate \{Cu$_2$(O$_2$CH)$_4$\}\{Cu(O$_2$CH)$_2$(2-methylpyridine)$_2$\}
is found recently which supports entanglement above room temperature \cite{Souza}.

In this article, a cluster of six spins with all-to-all 
two-spin AFM exchange interactions is 
considered which gives rise to quantum states with multipartite entanglement
those can withstand thermal agitations. 
The model has been solved exactly to obtain all analytic expressions of 
eigenvalues and eigenfunctions. 
Symmetry of each eigenstate is studied by exploiting the six-fold 
rotational invariance of the Hamiltonian. 
Bipartite entanglements have been characterized with the help of 
concurrence (CN) while the multipartite entanglements are studied by 
introducing 
several EWs.  
The model Hamiltonian is introduced in the Sec. \ref{model} 
along with the characterization of frustration embedded in it. 
The properties of thermal CN 
have been discussed in the section \ref{TC}. 
Detection of bipartite and multipartite entanglements in terms of 
EWs based on susceptibility, fidelity and internal energy  
is presented in the Sec. \ref{ews} while Sec. 
\ref{discussion} holds a comprehensive discussion on the results.  
\section{The SPIN-$\frac{1}{2}$ AFM $J_1$-$J_2$-$J_3$ HEISENBERG HEXAGON} 
\label{model}
\begin{figure}[h]
\begin{center}
\psfrag{1}{$\vec S_1$}
\psfrag{2}{$\vec S_2$}
\psfrag{3}{$\vec S_3$}
\psfrag{4}{$\vec S_4$}
\psfrag{5}{$\vec S_5$}
\psfrag{6}{$\vec S_6$}
\psfrag{J1}{$J_1$}
\psfrag{J2}{$J_2$}
\psfrag{J3}{$J_3$}
\psfrag{A}{(a)}
\psfrag{B}{(b)}
\psfrag{C}{(c)}
\psfrag{D}{(d)}
\psfrag{E}{(e)}
\psfrag{ps1}{$\Psi_{\rm RVB}$}
\psfrag{ps2}{$\Psi_{\rm RVB}'$}
\psfrag{i}{$i$}
\psfrag{e}{$\equiv$}
\psfrag{f}{$\frac{1}{\sqrt 2}$}
\psfrag{b}{$($}
\psfrag{a}{$)$}
\psfrag{u}{$\uparrow$}
\psfrag{d}{$\downarrow$}
\psfrag{m}{$-$}
\psfrag{j}{$j$}
\psfrag{t}{$\sqrt{\frac{2}{3}}$}
\psfrag{p}{$\frac{2}{\sqrt{3}}$}
\psfrag{al}{$\frac{1}{\alpha^\prime_5}$}
\psfrag{c}{$C^\prime_{51}$}
\psfrag{o}{$+$}
\psfrag{,}{,}
\psfrag{M}{\text{\scriptsize{M}}}
\psfrag{P}{\text{\scriptsize{P}}}
\psfrag{g}{$|\Gcal\rangle$}
\psfrag{L}{\text{\scriptsize{L}}}
\psfrag{cc}{\text{\scriptsize{C}}}
\psfrag{O}{\text{\scriptsize{O}}}
\psfrag{0.0}{\text{\scriptsize{0.0}}}
\psfrag{1.0}{\text{\scriptsize{1.0}}}
\psfrag{0.5}{\text{\scriptsize{0.5}}}
\psfrag{r1}{\text{\scriptsize{R$_1$}}}
\psfrag{r2}{\text{\scriptsize{R$_2$}}}
\psfrag{j3}{\text{\scriptsize{ $J_3/J_1$}}}
\psfrag{j2}{\text{\scriptsize{ $J_2/J_1$}}}
  \includegraphics[scale=0.37]{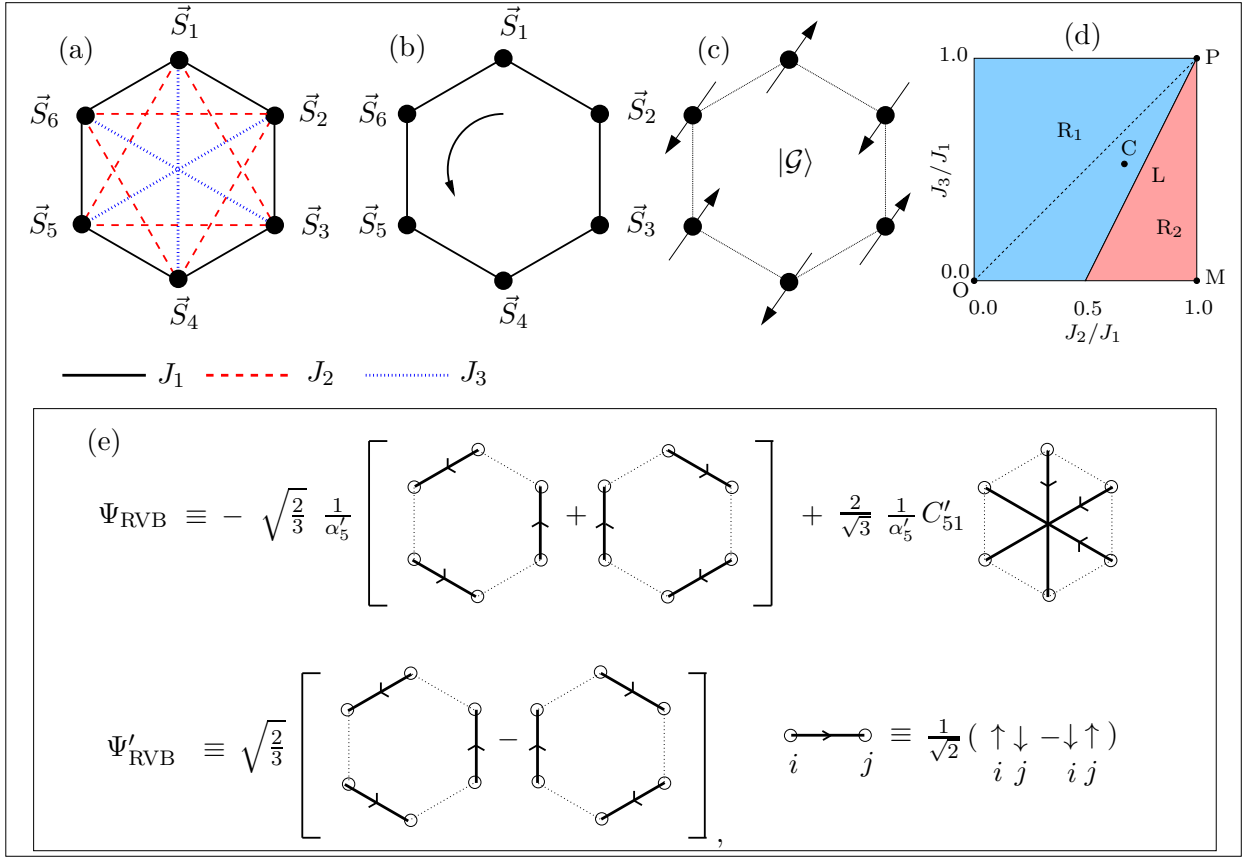}  
  \caption{(a) Geometrical view of the model, (b) rotation by $60^\circ$ keeps the system invariant, 
(c) classical ground state $|\Gcal\rangle$, (d) the $J_2$-$J_3$ parameter space 
and (e) quantum ground states, $\Psi_{\rm RVB}$ and $\Psi_{\rm RVB}'$. 
Expressions of $\alpha^\prime_5$ and $C^\prime_{51}$ are shown in the Appendix \ref{eigensystem}.}
  \label{fmodel}
\end{center}
\end{figure}

Spin-$\frac{1}{2}$ AFM Heisenberg Hamiltonian on the 
hexagonal cluster is defined by
\begin{eqnarray}
 H&=&H_{\rm NF}+H_{\rm F},  \label{ham} \\
H_{\rm NF}&=&J_1 \sum_{i=1}^6 \vec S_i \cdot \vec S_{i+1} 
+J_3\sum_{i=1,2,3}\vec S_i\cdot \vec S_{i+3},\nonumber\\
H_{\rm F}&=&J_2  \sum_{i=1}^6 \vec S_i \cdot \vec S_{i+2}, \;
\vec S_{i+6}=\vec S_{i}.\nonumber
\end{eqnarray}
$\vec S_i$ is the spin-1/2 operator at the position $i$.
In this model, every spin is interacting with all other spins 
via the AFM exchange interactions. 
As a result, three topologically different exchange couplings, say, 
nearest neighbor (NN), next nearest neighbor 
(NNN) and further neighbor (FN) or diagonal exchanges appear 
whose strengths are $J_1$, $J_2$ and $J_3$, respectively.  
Geometrical view of this spin model is shown in the Fig. \ref{fmodel}(a). 
Frustration appears in a magnetic system when all the AFM bonds 
are not energetically minimized in the classical ground state simultaneously. 
In this model, $J_2$ is frustrating, while 
$J_1$ and $J_3$ are non-frustrating. 
With this view, the total Hamiltonian, $H$, (Eq. \ref{ham}) is decomposed into 
two parts, non-frustrated ($H_{\rm NF}$) and frustrated ($H_{\rm F}$). 
Frustration does not appear in this system if $J_2$ is assumed 
negative (ferromagnetic). 
For $J_2< (J_1 + J_3/2)$, 
the classical ground state of this model is a doublet, where each state 
is connected to other by flipping the spins in every site.  
One of such state, $|\mathcal{G}\rangle$, is shown in the Fig. \ref{fmodel}(c), 
in which the adjacent spins are antiparallel.  
In this case, energy minimization for both $H_{\rm NF}$ and $H_{\rm F}$ cannot be taken 
place simultaneously in the ground state. 
Energy minimization of an AFM bond occurs when the spin alignments around this bond 
is antiparallel. As a result, energy corresponding to $H_{\rm NF}$ with respect to 
the ground state $|\Gcal\rangle$ is minimized but 
that of $H_{\rm F}$ is maximized with respect to 
the same $|\Gcal\rangle$. The frustration of this model can be 
characterized by using the quantity, frustration degree ($\mathcal F$) 
which is defined as \cite{Sen},   
\be
\mathcal F={\rm avg}\frac{\langle \Gcal |H_{\rm F}|\Gcal\rangle}
{|\langle \Gcal|H_{\rm NF}|\Gcal \rangle|},
\ee
where ``avg'' denotes the averaging over all possible ground states.
In this model, $\mathcal F=J_2/(J_1+J_3/2)$. 
For the frustrated system  $\mathcal F>0$, while it 
is non-frustrated when $\mathcal F\leq 0$. 
The higher value of $\mathcal F$ corresponds to the stronger frustration. 
The variation of  $\mathcal F$ in the $J_2$-$J_3$ parameter space 
is shown in the Fig. \ref{fourplot1} (a). 
$\mathcal F$ is found to increase (decrease) with the increase of $J_2$ ($J_3$).  
The maximum value of 
$\mathcal F$ is unity which appears at the point $J_3=0$ over the line $J_2=J_1$ in 
the parameter space. 
This particular point is labeled by the letter M in the 
parameter space (Fig. \ref{fmodel} (d)). 
Therefore, the system is maximally frustrated at the point M. 
On the other hand, the minimum value of $\mathcal F$ is zero for this AFM model 
which is observed 
over the line $J_2=0$, where the system is said to be non-frustrated.
Thus, effects of magnetic frustration on the entanglement properties 
can be studied with this model. 
The Hamiltonian, Eq. \ref{ham}, commutes with total spin operator, $S_{\rm T}$, 
as well as the $z$-component of the total spin, $S^z_{\rm T}$.
 As a result, the Hamiltonian may be spanned in the 
different subspaces of $S^z_{\rm T}$ to obtain analytic expressions of eigenvalues and 
eigenfunctions.  
The exact analytic expressions of all 64  
eigenstates ($\Psi_n,\;n=1,2,3,\cdots, 64$) and corresponding energy eigenvalues ($E_n$)
are available in the Appendix \ref{eigensystem}. 
Those states essentially comprise to five singlets 
($S_{\rm T}=0$), nine triplets ($S_{\rm T}=1$), 
five quintets ($S_{\rm T}=2$) and one septet ($S_{\rm T}=3$). 
Five distinct singlets are denoted by the eigenstates $\Psi_{38}$, $\Psi_{39}$, 
$\Psi_{40}$, $\Psi_{41}$ and $\Psi_{42}$ in the Appendix \ref{eigensystem}.
Among the five, two singlets, $\Psi_{42}$ and $\Psi_{38}$ can be expressed  
by two distinct combinations of 
dimer states which are known as resonating valence bond (RVB) states. 
Those two particular singlets are defined by $\Psi_{\rm RVB}$ and  
$\Psi_{\rm RVB}'$. 
The arrangements of dimer states in $\Psi_{\rm RVB}$ ($\Psi_{42}$) and  
$\Psi_{\rm RVB}'$ ($\Psi_{38}$) are shown in the Fig. \ref{fmodel} (e).  
Ground state is always a total spin singlet. 
All the five singlets participate in four different manners  
to constitute the ground state in the whole parameter space. 
Thus, depending on the combinations of singlets in the ground states,  
$J_2$-$J_3$ parameter space is decomposed into four segments. 
$\Psi_{\rm RVB}$ and  
$\Psi_{\rm RVB}'$ are the ground states (non-degenerate) in the regions, 
R$_1$ ($J_1+J_3>2J_2$) and 
R$_2$ ($J_1+J_3<2J_2$), respectively. $\Psi_{\rm RVB}$ and  
$\Psi_{\rm RVB}'$ form the doubly degenerate ground state over the line, L 
($J_1+ J_3 = 2 J_2$), junction of the two regions, R$_1$ and R$_2$. 
And all the five singlets constitute the ground state 
(five-fold  degenerate) at the point P ($J_1= J_3 = J_2$). 
Positions of R$_1$, R$_2$, L and P on the parameter 
space are shown in the Fig. \ref{fmodel} (d). The area of R$_1$ 
is three times larger than that of R$_2$. 
A first order QPT occurs 
across the line L as well as at the point P,   
where the ground state cross over takes place. 

In addition, the Hamiltonian possesses another useful symmetry,  
where it is invariant under the rotation by 
$60^\circ$, (Fig. \ref{fmodel}(b)).  
For the counter clockwise rotation by $60^\circ$, 
a rotational operator, $\hat{R}$, can be defined as  
$\hat{R}\bra {S_1S_2S_3S_4S_5S_6}=\bra{S_2S_3S_4S_5S_6S_1}$, where 
$\bra{S_1S_2S_3S_4S_5S_6}=\bra{S_1^z}\otimes\bra{S_2^z}\otimes\bra{S_3^z}
\otimes\bra{S_4^z}\otimes\bra{S_5^z}\otimes\bra{S_6^z}$, in which 
$\bra{S_i^z}$ is the spin state at site $i$.
So, $\hat{R}^{(n)}$ be the successive $\hat{R}$ operation by 
$n$ times, such that $\hat{R}^{(6)}$ is the identity 
operation which leaves any state 
unchanged. Each eigenstate ($\Psi$) of the Hamiltonian 
has some definite rotational property, 
which can be characterized in terms of an eigenvalue equation, like 
$\hat{R}^{(n)}|\Psi\rangle=\lambda_r|\Psi\rangle$, where  
$\lambda_r$'s are the eigenvalues of the rotational operator $\hat{R}^{(n)}$. 
$\lambda_r$ can assume the value either $+1$ or $-1$ 
for the minimum number ($p$) of 
 $\hat{R}$ operations on a definite state. 
Obviously, for the same state $\lambda_r$ is always $+1$ for $2p$ number of 
 $\hat{R}$ operations. 
The states with $\lambda_r=+1$ for $p$ number of 
 $\hat{R}$ operations have even parity (symmetric) 
while those with $\lambda_r=-1$ have odd parity (antisymmetric). 
It is found that, every eigenstate has definite values 
of both $p$ and $\lambda_r$, and thus has definite parity. 
36 states have even parity while the remaining 28 states have odd parity. 
Values of $p$ and $\lambda_r$ for all eigenstates are shown   
in the Tab. I. It is observed that 
$p$ takes up  either 1 or 3 and never 
takes up 2, 4 and 5.  For $\Psi_{\rm RVB}$, $\lambda_r=-1$ and $p=1$, while,  
for $\Psi_{\rm RVB}'$, $\lambda_r=1$ and $p=1$. Thus, 
$\Psi_{\rm RVB}'$ does not change sign 
under any number of $\hat{R}$ operations, while $\Psi_{\rm RVB}$ changes sign 
for odd numbers of $\hat{R}$ operations. So, $\Psi_{\rm RVB}$ is 
antisymmetric, whereas, $\Psi_{\rm RVB}'$ is symmetric under the rotation by 
$60^\circ$.  
\section{Thermal Concurrence} 
\label{TC}
\begin{figure}[h]
\begin{center}
\psfrag{M}{\text{\footnotesize{M}}}
\psfrag{P}{\text{\footnotesize{P}}}
\psfrag{F}{\text{\footnotesize{$\Fcal$}}}
\psfrag{L}{\text{\footnotesize{L}}}
\psfrag{R1}{\text{\footnotesize{R$_1$}}}
\psfrag{R2}{\text{\footnotesize{R$_2$}}}
\psfrag{J3}{\text{\footnotesize{ $J_3/J_1$}}}
\psfrag{J2}{\text{\footnotesize{ $J_2/J_1$}}}
\psfrag{a}{(a)}
\psfrag{b}{(b)}
\psfrag{c}{(c)}
\psfrag{d}{(d)}
\psfrag{kT}{\text{\footnotesize{$k_{\rm B}T/J_1$}}}
\psfrag{CN2}{\footnotesize CN$_{\rm NN}$}
\psfrag{CN3}{\footnotesize CN$_{\rm FN}$}
\psfrag{CN4}{\footnotesize TCN$_{\rm NN}$}
\includegraphics[scale=1.10]{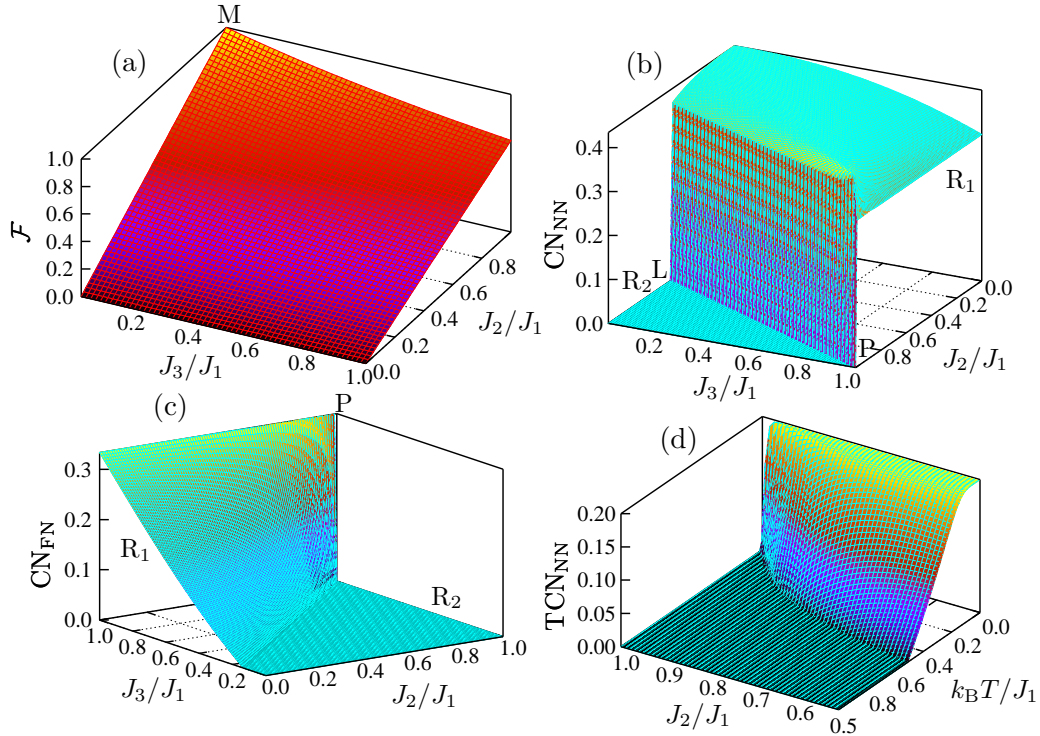}
\caption{Variations of frustration degree, $\Fcal$, (a) 
 CN$_{\rm NN}$ (b) and CN$_{\rm FN}$ (c) with respect to $J_2/J_1$ and $J_3/J_1$.
(d) Variation of TCN$_{\rm NN}$ with respect to $J_2/J_1$ 
and $k_{\rm B}T/J_1$ over the line L including the point P. 
Orientations of $J_2/J_1$ and $J_3/J_1$ axes are made 
different for different figures to have more clarity.}
\label{fourplot1}
\end{center}
\end{figure}

For the Heisenberg hexagon, thermal state density matrix has been written down as 
\be
\rho(T)=\frac{1}{Z}\sum\limits_{n=1}^{64}e^{-\beta E_n}\rho^n;\quad \rho^n=\bra{\Psi_n}\ket{\Psi_n}, 
\ee
 where $Z$ is the partition function of the system.
$\beta^{-1}=k_{\rm B}T$, where $k_{\rm B}$ and $T$ are the 
Boltzmann constant and temperature, respectively. 
Eigenvalues, $E_n$ and the corresponding eigenstates, $\Psi_n$ are 
shown in the Appendix \ref{eigensystem}.  
Similarly, the reduced thermal state density matrix $\rho_{ij}(T)$ 
can be written as, 
\be
\rho_{ij}(T)=\frac{1}{Z}\sum\limits_{n=1}^{64}e^{-\beta E_n}\rho_{ij}^n,   
\ee
where the reduced density matrix, $\rho_{ij}^n$ is obtained 
from $\rho^n$  by tracing out
the remaining four spin degrees of freedom, those are 
not located at the sites $i$ and $j$. 
CN is one of the simplest measure to quantify the entanglement 
between two qubits when they sit at two different sites in the surrounding of 
other interacting spins and that can be derived from the 
expression of $\rho_{ij}(T)$. 
At $T=0$, $\rho_{ij}(T)$ reduces to $\rho^{\rm G}_{ij}$, where 
$\rho^{\rm G}=\bra{\Psi_{\rm G}}\ket{\Psi_{\rm G}}$ and $\Psi_{\rm G}$ 
is the ground state. $\Psi_{\rm G}$ becomes equal to 
$\Psi_{42}$ and $\Psi_{38}$ for the regions R$_1$ and R$_2$, respectively. 
On the line L, $\rho^{\rm G}=(\bra{\Psi_{38}}\ket{\Psi_{38}}+\bra{\Psi_{42}}\ket{\Psi_{42}})/2$. 
Similarly, at P $\rho^{\rm G}=(\bra{\Psi_{38}}\ket{\Psi_{38}}+\bra{\Psi_{39}}\ket{\Psi_{39}}+
\bra{\Psi_{40}}\ket{\Psi_{40}}+\bra{\Psi_{41}}\ket{\Psi_{41}}+\bra{\Psi_{42}}\ket{\Psi_{42}} )/5$. 
Depending on the 
positions of the sites $i$ and $j$, only three 
different types of  $\rho^{\rm G}_{ij}$ can be constructed.  
They are $\rho^{\rm G}_{\rm NN}$, $\rho^{\rm G}_{\rm NNN}$ and $\rho^{\rm G}_{\rm FN}$,
when the sites  $i$ and $j$ are NN, NNN and FN, respectively. 
For example, there is six distinct pairs of 
NN sites for different values of 
$i$ and $j$ (\{$ij$\}), say, \{12\}, \{23\}, \{34\}, \{45\}, \{56\} and \{61\}. 
$\rho^{\rm G}_{ij}$ is same for all these six NN pairs by virtue 
of the rotational symmetry of hexagon. So, they 
are abbreviated as $\rho^{\rm G}_{\rm NN}$. 
The similar argument holds true for other combinations, NNN and FN. 
NNN corresponds to six distinct pairs while FN corresponds to only three pairs.
The general form of two-qubit $\rho^{\rm G}_{ij}$ in the space of $S^z$ diagonal basis states,    
$\{\bra{\uparrow \uparrow},\,\bra{\uparrow \downarrow} ,\,
 \bra{\downarrow\uparrow},\,\bra{\downarrow\downarrow}\}$, looks like, 
  \begin{eqnarray}
  \rho^{\rm G}_{ij}=
\left[ 
 { \begin{array}{cccc}
  a & 0 & 0 & f \\
  0 & b_1 & z & 0 \\
  0 & z^* & b_2 & 0 \\
  f^* & 0 & 0 & d
  \end{array}}
\right].
\label{rho}
\end{eqnarray}
By expressing  $\rho^{\rm G}_{ij}$ in this form 
one can define the spin reversed reduced density matrix as, 
$\overline{\rho^{\rm G}_{ij}}=(\sigma_y\otimes\sigma_y)(\rho^{\rm G})^{{}^*}_{ij}
(\sigma_y\otimes\sigma_y)$, 
where $\sigma_y$ is the Pauli matrix. Then concurrence between the sites 
$i$ and $j$ (CN$_{ij}$) is given by 
CN$_{ij}={\rm max}\{\lambda_1-\lambda_2-\lambda_3-\lambda_4,0\}$, 
where $\lambda_i$s are the square roots of the eigenvalues of the non-Hermitian matrix 
$\rho^{\rm G}_{ij}\,\overline{\rho^{\rm G}_{ij}}$, in descending order \cite{Wooters_prl80}. 
Since $S^z_{\rm T}$ is the good quantum number,   
the element $f$  in $\rho^{\rm G}_{ij}$ (Eq. \ref{rho}) vanishes. 
As a result, the expression of concurrence looks simpler, which 
is given by \cite{Wooters_pra63}
\begin{eqnarray}
 \textrm{CN}_{ij}=2 \;\textrm{max}\left(0,|z|-\sqrt{ad}\right). 
 \label{cn}
 \end{eqnarray}
CN$_{ij}$ measures the pairwise entanglement between two spins at sites $i$ and $j$, 
which varies from CN$_{ij}=0$ for a separable state to 
CN$_{ij}=1$ for a maximally entangled state.  
Variations of CN$_{\rm NN}$ and CN$_{\rm FN}$ for four different 
locations in the parameter space are shown in Fig. \ref{fourplot1} (b) and (c), respectively. 
CN$_{\rm NNN}$ is zero everywhere which means that 
 concurrence between NNN sites does not survive. 
CN$_{\rm NN}$ is found to obey the relation CN$_{\rm NN}=-\frac{1}{2}[4E_{\rm G}/N+1]$, 
for $J_2=J_3=0$ where $E_{\rm G}$ is the ground state energy of 
$S=\frac{1}{2}$ AFM Heisenberg chain with $N$ sites and periodic boundary 
condition \cite{Wooters_pra63}. Similar types of relations for CN$_{\rm NNN}$ and 
CN$_{\rm FN}$ are not found. CN$_{\rm NN}=0.434$ over the line $J_2=J_3$ 
which is also maximum. This particular line 
is marked by the dashed line OP 
in the parameter space (Fig. \ref{fmodel} (d)). CN$_{\rm NN}$ vanishes in the 
region R$_2$. 
CN$_{\rm NN}$ suffers a jump over the line L, 
which is the signature of a first-order QPT. 
In R$_1$, for fixed value of the frustrating bond ($J_2$), 
CN$_{\rm NN}$ increases with $J_3$ 
up to the line OP, where it acquires the maximum value.
With further increase of $J_3$, it begins to decrease.
On the other hand, 
CN$_{\rm FN}$ is zero throughout the region R$_2$ in addition to the 
portion of R$_1$ where  $J_3\leq(0.87J_2+0.14)$.  
In the region R$_1$, for fixed $J_2$, CN$_{\rm FN}$ 
increases with the increase of $J_3$ but for fixed $J_3$, it 
decreases with increasing $J_2$.
The maximum value of CN$_{\rm FN}$ is observed over the line 
$J_3=J_1$ barring the point P.  
There is no effect of frustration on 
CN$_{\rm FN}$ in the locations R$_2$ and L. 
On the other hand, they tend to decrease with the increase of $J_2$ in R$_1$. 

\begin{center}
\begin{figure}[h]
\psfrag{a}{\text{\footnotesize{(a)}}}
\psfrag{b}{\text{\footnotesize{(b)}}}
\psfrag{J2}{\text{\footnotesize{  $J_2/J_1$}}}
\psfrag{J3}{\text{\footnotesize{ $J_3/J_1$}}}
\psfrag{T1}{\text{\footnotesize{$k_{\rm B}T^{\rm NN}_{\rm c}/J_1$}}}
\psfrag{T2}{\text{\footnotesize{$k_{\rm B}T^{\rm FN}_{\rm c}/J_1$}}}
\includegraphics[scale=1.0]{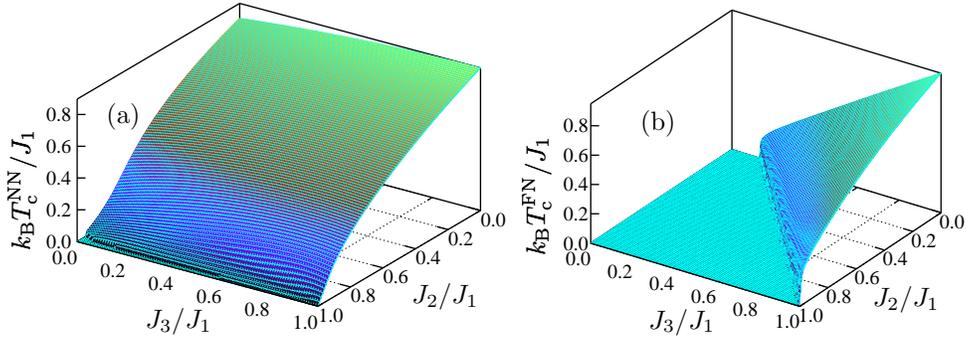}
  \caption{Variation of (a) $k_{\rm B}T^{\rm NN}_{\rm c}/J_1$ and 
(b) $k_{\rm B}T^{\rm FN}_{\rm c}/J_1$ with respect to $J_2/J_1$ and  $J_3/J_1$.}
\label{twoplot}
\end{figure}
 \end{center}

The thermal state concurrence (TCN) 
has been derived from $\rho_{ij}(T)$ by using Eqs. (\ref{cn}). 
The variations of TCN$_{\rm NN}$ with respect to 
$k_{\rm B}T/J_1$ for the line L including the point P has 
 been displayed in Fig. \ref{fourplot1} (d).
TCN decreases with temperature  and 
exactly vanishes at the critical temperature $T^{ij}_{\rm c}$. 
Non-zero values for $T^{\rm NN}_{\rm c}$ and 
$T^{\rm FN}_{\rm c}$ have been observed while $T^{\rm NNN}_{\rm c}$ 
is always zero. Variations of $T^{\rm NN}_{\rm c}$ and 
$T^{\rm FN}_{\rm c}$ have been shown in 
 Fig. \ref{twoplot} (a) and (b), respectively. 
For a fixed $J_3$, both $T^{\rm NN}_{\rm c}$ and $T^{\rm FN}_{\rm c}$ decrease 
very fast with $J_2$ whereas for fixed $J_2$, they both increase slowly with $J_3$. 
The variations of  both $T^{\rm NN}_{\rm c}$ and  $T^{\rm FN}_{\rm c}$ with respect to 
$J_2$ indicate that frustration opposes the bipartite entanglement in this system.

\section{entanglement witnesses: susceptibility, fidelity and internal energy}
\label{ews}
\begin{figure}[h]
\psfrag{T}{$k_{\rm B}T/J_1$}
\psfrag{F} {$F$}
\psfrag{cn} {TCN$_{\rm NN}$}
\psfrag{a}{(a)}
\psfrag{b}{(b)}
\psfrag{x2}{Eq.\ref{eq}}
\psfrag{x1}{Eq.\ref{chi}}
\psfrag{t1}{$T_{\rm c}\approx 1.43J_1/K_B$}
\psfrag{t2}{$T_{\rm c}^{\rm NN}\approx 0.80 J_1/k_{\rm B}$}
\psfrag{chi}{$\chi/(g^2\mu^2_{\rm B}/J_1)$}
\psfrag{fidelity}{TCN$_{\rm NN}$, $F$}
\centerline{\includegraphics[scale=0.60]{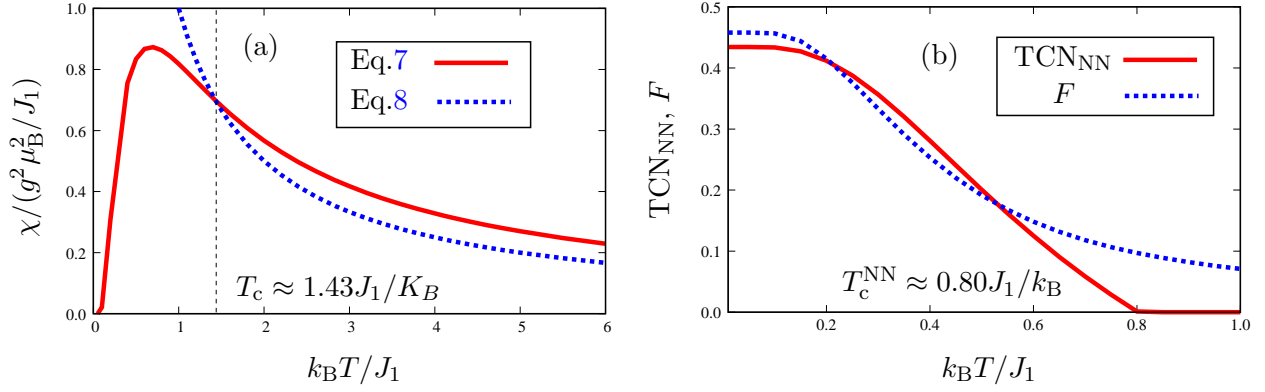}}
  \caption{(a) Variations of $\chi/(g^2\mu^2_{\rm B}/J_1)$ against  $k_{\rm B}T/J_1$ following 
the Eqs. \ref{chi} and \ref{eq}. 
(b) Variation of TCN$_{\rm NN}$ and $F$ against $k_{\rm B}T/J_1$. }
\label{t1cx}
\end{figure}

\begin{figure}[h]
\begin{center}
\psfrag{M}{\text{\footnotesize{M}}}
\psfrag{P}{\text{\footnotesize{P}}}
\psfrag{F}{\text{\footnotesize{$\Fcal$}}}
\psfrag{L}{\text{\footnotesize{L}}}
\psfrag{R1}{\text{\footnotesize{R$_1$}}}
\psfrag{R2}{\text{\footnotesize{R$_2$}}}
\psfrag{J3}{\text{\footnotesize{ $J_3/J_1$}}}
\psfrag{J2}{\text{\footnotesize{ $J_2/J_1$}}}
\psfrag{a}{(a)}
\psfrag{b}{(b)}
\psfrag{c}{(c)}
\psfrag{d}{(d)}
\psfrag{Esep}{\footnotesize{$E_{\rm sep}/J_1$}}
\psfrag{EGTs}{\text{\tiny{ $k_{\rm B}T_E/J_1$}}}
\psfrag{F}{\footnotesize {$F$}}
\psfrag{kT}{\text{\footnotesize{$k_{\rm B}T/J_1$}}}
\psfrag{CN3}{\footnotesize CN$_{\rm FN}$}
\psfrag{CN4}{\footnotesize TCN$_{\rm NN}$}
\psfrag{T1}{\footnotesize{$k_{\rm B}T_{\rm c}/J_1$}}
\psfrag{EGT}{\footnotesize {$k_{\rm B}T_{\rm E}/J_1$}}
\includegraphics[scale=1.10]{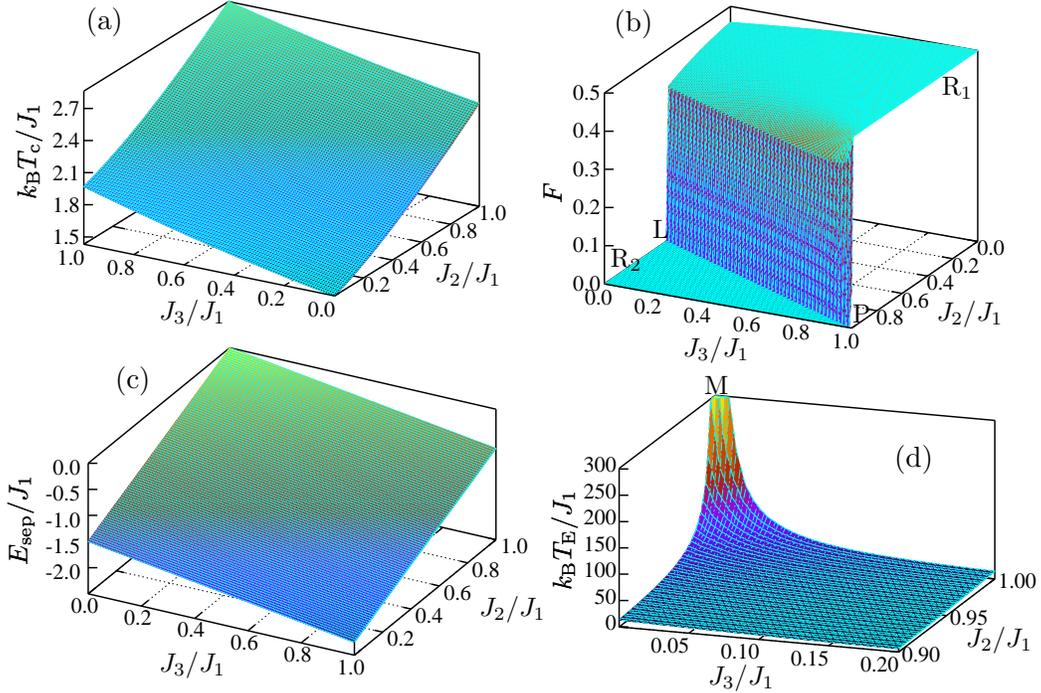}
\caption{Variation of $k_{\rm B}T_{\rm c}/J_1$ (a), $F$ (b), 
  $E_{\rm sep}/J_1$ (c) and  $k_{\rm B}T_{\rm E}/J_1$ (d) 
with respect to $J_2/J_1$ and $J_3/J_1$.}
\label{fourplot2}
\end{center}
\end{figure}

In 1996, Horodecki {\em et. al.} formulate the necessary and sufficient 
conditions for separability of a bipartite system \cite{Horodeckis}. This formulation leads to 
the existence of a particular EW which is essentially a measure of violation of Bell 
inequality \cite{Terhal}. 
For a magnetic system, it has been shown that magnetic susceptibility can serve 
as an EW which can be applied without complete knowledge of the Hamiltonian \cite{Wiesniak}.
 For an isolated  $N$-spin cluster, which is SU(2) invariant and translationally symmetric, 
the condition of untangled states has been put forward in 
term of an inequality \cite{Brukner}. 
For the Heisenberg Hamiltonian, which is isotropic in the spin space, 
the magnetic susceptibility along a particular direction, $\alpha$, ($\alpha=x,y,z$) is given by
 \[\chi_{\alpha}=\frac{\left(g\mu_{\rm B}\right)^2}{k_{\rm B}T}
\left(\langle M^2_\alpha\rangle-\langle M_\alpha\rangle ^2\right),\]
 where $M_\alpha=\sum\limits_{i=1}^{N} S^i_\alpha$ is the magnetization along the 
direction $\alpha$, 
 $g$ is the g-factor and $\mu_B$ is the Bohr magneton. Thus, 
 \[\chi_\alpha=\frac{\left(g\mu_{\rm B}\right)^2}{k_{\rm B}T}
\left(\sum\limits_{i,j=1}^{N}\langle S^i_\alpha S^j_\alpha\rangle-
 \langle\sum\limits_{i=1}^{N} S^i_\alpha\rangle^2\right).\]
 Since the Hamiltonian is isotropic in the spin space,  
 $\chi=\chi_x=\chi_y=\chi_z$, or, 
 $\chi=\frac{1}{3}\left(\chi_x+\chi_y+\chi_z\right)$, and 
$\langle\sum\limits_{i=1}^{N} S^i_\alpha\rangle=0$, $\chi$ can be expressed as 
 \begin{eqnarray}
  \chi=\frac{\left(g\mu_{\rm B}\right)^2}{k_{\rm B}T}
\left(\frac{N}{4}+\frac{2}{3}\sum\limits_{i<j}\langle \vec S_i. \vec S_j\rangle\right).
\label{chi}
 \end{eqnarray}
The second term in the expression of $\chi$, {\em i. e.}, 
the sum within the expectation value in Eq. \ref{chi},  
acts as the all-to-all spin interaction term. 
Alternately, in this particular case, this sum is equivalent to the 
Hamiltonian (Eq. \ref{ham}), at the point P when $J_1=1$, say, $H_{\rm P}$. As a result, 
$\sum\limits_{i<j}\langle \vec S_i. \vec S_j\rangle=\langle H_{\rm P}\rangle$ 
corresponds to the ground state energy at the point P for $J_1=1$. 
Due to AFM spin interaction the ground state expectation value of 
$H_{\rm P}$ is always negative. So, $\langle H_{\rm P}\rangle$ 
makes a negative contribution to $\chi$. 
And the maximum negative value of $\langle H_{\rm P}\rangle$ 
is equal to the ground state energy of $H_{\rm P}$ itself. 
It has been discussed in the next section that minimum energy of the 
separable states is negative and equivalent to the ground-state 
energy of the corresponding classical Hamiltonian. 
For $N=6$, $\langle H_{\rm P}\rangle=-3/4$.
For any general separable states,  $\langle H_{\rm P}\rangle$ always make 
lesser contribution to $\chi$ in comparison to 
the  separable state of minimum energy. 
Therefore, for a single cluster of $N=6$ spin the condition 
of untangled states has been given 
 by the inequality
  \begin{eqnarray}
  \chi\geq\frac{(g\mu_B)^2}{k_BT}. 
  \label{eq}
 \end{eqnarray}
Curves describing the variation of $\chi/(g^2\mu^2_{\rm B}J_1)$ against $k_{\rm B}T/J_1$
arising from the above equality, Eq. \ref{eq} and the same variation resulting 
from  Eq. \ref{chi} intersect at a critical temperature, $T_{\rm c}$, below 
which the system is entangled. Thus, $\chi$, (Eq. \ref{chi}) acts as an EW. 
The variations of $\chi/(g^2\mu^2_{\rm B}J_1)$ against $k_{\rm B}T/J_1$
representing  Eqs. \ref{chi} and \ref{eq} have been 
shown in Fig. \ref{t1cx} (a).  
Eq. \ref{chi} has been evaluated for $N=6$ where only NN 
interactions are considered. 
Two curves intersects at $T_{\rm c}\approx 1.43J_1/K_B$. 
The variation of TCN$_{\rm NN}$ with respect to $k_{\rm B}T/J_1$ 
has been shown in Fig. \ref{t1cx} (b), where only NN interactions are considered.
This variation indicates that $T_{\rm c}^{\rm NN}\approx 0.802 J_1/k_{\rm B}$, 
where $T_{\rm c}^{\rm NN}$ is that critical temperature beyond which the  
bipartite entanglement does not exist. 
By comparing the values of $T_{\rm c}$ and $T_{\rm c}^{\rm NN}$, 
it is obvious that only multipartite entanglement is present
in the system in the intermediate temperature range $T_{\rm c}^{\rm NN}<T<T_{\rm c}$.  
Thus, below $T_{\rm c}^{\rm NN}$, both bipartite and multipartite entanglements are 
present while they vanish above $T_{\rm c}$.   
The variation of $k_{\rm B}T_{\rm c}/J_1$ for the AFM 
Heisenberg hexagon has been shown in Fig. \ref{fourplot2} (a). 
$T_{\rm c}$ has the maximum value at the point P when 
$J_1=J_2=J_3$, {\em i. e.}, where all-to-all interactions of equal strength are present. 
The minimum value of $T_{\rm c}$ appears when $J_2=J_3=0$, 
{\em i. e.}, where only NN interactions are present. 
With the increase of both $J_2$ and $J_3$, $T_{\rm c}$ increases steadily. 
But the rate of increase of $T_{\rm c}$ with respect to $J_2$ 
is more than that of $J_3$, which means that 
frustration enhances the multipartite entanglement in the system. 

In order to investigate the presence of six-qubit entanglement in the 
AFM Heisenberg hexagon, the state preparation fidelity, $F$ is 
defined as, $ F(\rho)=\ket{\Psi_{GHZ}}\rho(T)\bra{\Psi_{GHZ}}$, 
 where $\bra{\Psi_{GHZ}}=\frac{1}{\sqrt{2}}\left(\bra{\uparrow\downarrow
\uparrow\downarrow\uparrow\downarrow}
 -\bra{\downarrow\uparrow\downarrow\uparrow\downarrow\uparrow}\right)$ 
is the six-spin Greenberger-Horne-Zeilinger (GHZ) state \cite{Wang}.
The sufficient condition for the presence of six-particle entanglement 
in this six-qubit system is given by the inequality, 
$ F(\rho)>\frac{1}{2}$ \cite{Sackett,Bennett}. 
For the hexagonal system with $J_2=J_3=0$, variation of $F(\rho)$ against 
$k_{\rm B}T/J_1$ has been shown in Fig. \ref{t1cx} (b). 
The variation of ground state fidelity $F$ in the  
parameter space is shown in Fig. \ref{fourplot2} (b). The value of $F$ 
is fixed over the line OP and that value of $F$ is $0.458$. 
The maximum value of $F$ at zero temperature is $1/2$ which is observed 
over another line  $J_3=J_1$ except the point P. 
$F$, however, vanishes over the entire region R$_2$. 
The value of $F$ just over the line L is fixed, and it 
suffers a sudden jump, which is the manifestation of QPT.
$F$ decreases with the increase of $T$ throughout the parameter space. 
Since $F(\rho)\leq\frac{1}{2}$, the six-spin entanglement is 
absent in the ground as well as the thermal states 
at all temperatures. 
On the other hand, for $S=\frac{1}{2}$ AFM 
Heisenberg tetramer with NN interaction, $F=\frac{2}{3}>\frac{1}{2}$, 
which indicates the presence of 
four-particle entanglement in ground state \cite{Wang,Bose}.
In general,  $F$ increases with $J_3$ for fixed $J_2$ and 
decrease with $J_2$ for fixed $J_3$. Therefore, frustration opposes the 
six-spin entanglement in this case. 

Another kind of detection for EW has been introduced 
by Dowling and others based on a comparison between 
the internal energy ($U(T)$) at finite temperature, $T$, and 
the minimum separable energy ($E_{\rm sep}$) \cite{Dowling}. 
The entanglement gap energy, $G_{\rm E}$ is defined by 
$G_{\rm E}(T)=E_{\rm sep}-U(T)$, at non-zero 
temperature  while that at zero temperature is given by 
$G_{\rm E}(0)=E_{\rm sep}-E_{\rm G}$, where $E_{\rm G}$ is the ground state energy.  
$U(T)$ is given by $U(T)=-\frac{1}{Z}\frac{\partial Z}{\partial \beta}$.
The multipartite entanglement would be present in the 
system at non-zero temperature, whenever $G_{\rm E}(T)>0$. 
With the increase of $T$, $G_{\rm E}(T)$ decreases 
since $U(T)$ increases with $T$. Obviously, there would be a limiting value 
of  $T$ above which $G_{\rm E}(T)<0$. 
This critical value of temperature, known as the entanglement 
gap temperature ($T_{\rm E}$) is define by, $U(T_{\rm E})=E_{\rm sep}$. 
Therefore, below $T_{\rm E}$  multipartite entanglement is non-zero. 
Thus a thermal state is entangled if $T<T_{\rm E}$.
To formalize this analysis, an EW, $Z_{\rm EW}$, a Hermitian operator 
is introduced such that Tr[$Z_{\rm EW}\rho_{\rm ent}$]$<0$, when 
there exists an entangled state, $\rho_{\rm ent}$. It is noted that 
$Z_{\rm EW}$ witnesses multipartite entanglement in $\rho_{\rm ent}$.
Therefore, positive entanglement gap, $G_{\rm E}(T)>0$, defines the 
EW by the equation $Z_{\rm EW}=H-IE_{\rm sep}$, 
where $I$ is the identity matrix on the Hilbert space. 
Hence, Tr[$Z_{\rm EW}\rho_{\rm sep}$]$=$Tr[$H\rho_{\rm sep}$]$-E_{\rm sep}\geq 0$, when 
 $\rho_{\rm sep}$ is any separable state while $E_{sep}$ is the lowest possible 
energy for a separable state. On the other hand, for the ground state, 
 $\rho_{\rm G}$, Tr[$Z_{\rm EW}\rho_{\rm G}$]$=E_{\rm G}-E_{\rm sep}< 0$ at $T=0$. 
Thus, $Z_{\rm EW}$ serves as an EW. 

Generally variational methods are being employed to find the 
lowest possible energy for a separable state of spin chains.
Otherwise, it has been noted that for AFM Heisenberg spin cluster 
with all-to-all couplings of same strengths,  
a minimum energy separable state is given by that classical 
spin configuration where the total spin vector is zero \cite{Dowling}.  
In order to find the separable state with minimum energy 
in this case, we 
introduce the most general form of separable state, like,  
$|\psi_{\rm sep}\rangle=\prod_j|S_j\rangle$, $j=1,2,3,\cdots, 6$, 
where $|S_j\rangle=\cos{\theta_j}|\uparrow\rangle +e^{i\phi_j}
\sin{\theta_j}|\downarrow\rangle$, 
$0\leq\theta_j\leq \pi$, and $0\leq\phi_j\leq 2\pi$. $E_{\rm sep}$ 
is obtained by minimizing $\langle \psi_{\rm sep}|H|\psi_{\rm sep}\rangle$ 
with respect to both $\theta_j$ and $\phi_j$. By using simplex minimizing 
procedure \cite{Nelder_Mead}, $E_{\rm sep}$ 
is found to equal to $ -\frac{3}{2} (J_1 - J_2) - \frac{3}{4} J_3$,  
which essentially corresponds to $\theta_j=\frac{2\pi j}{6}$ and $\phi_j=0$. 
The symmetry in the Hamiltonian is responsible for the 
symmetric solutions. 
The solutions always correspond to the classical spin configuration  
with the total spin vector is zero, 
although the condition of all-to-all couplings of same strength is 
mostly violated except the point P. 
The variations of $E_{\rm sep}/J_1$ and $k_{\rm B}T_{\rm E}/J_1$ 
are shown in Figs. \ref{fourplot2} (c) and (d), respectively. 
Usually $E_{\rm sep}$ is negative everywhere except 
the extreme point, M, over the line $J_1=J_2$, where $E_{\rm sep}$ becomes zero. 
At the point M, the value of frustration degree, $\Fcal$ is the maximum and 
$T_{\rm E}$ tends to $\infty$, which is shown in the Fig. \ref{fourplot2} (d). 
The value of $k_{\rm B}T_{\rm c}/J_1$ at this point is 2.395.  
The bipartite entanglement vanishes over the same line including this point. 
Therefore, at this point multipartite entanglement survives 
at all temperatures in the absence of bipartite entanglement. 
Besides this particular point, M, entangled states are found to exist at high 
temperatures in the vicinity of the point. 
Thus, the entanglement in quantum states in this particular region 
exhibits a robustness to the thermal noise. 
It appears from the expression of $E_{\rm sep}$ that positive contribution 
to $E_{\rm sep}$ only comes from the NNN frustrating bond, $J_2$. 
So, in the absence of NNN bond, $E_{\rm sep}$ is always negative which gives rise to 
very low $T_{\rm E}$. Therefore, the presence of frustration leads 
to the high values of $T_{\rm E}$. 
This observation shows that the frustration 
induces the multipartite entanglement in this spin 
cluster in such a manner that it does sustain against the thermal agitation.
On the other extreme point P over the same line $J_1=J_2$,  
it is found that $T_{\rm E}=T_{\rm c}= 2.862 J_1/K_{\rm B}$. 
Since the bipartite entanglement vanishes over this line 
only multipartite entanglement survives in the system at P for $0<T<T_{\rm E}$. 
The equality between $T_{\rm E}$ and $T_{\rm c}$ results from the fact that 
 at this point the effective spin interactions are defined 
on a non-bipartite graph or lattice for which EW based on thermal energy 
detects only the multipartite entanglement.
Now consider another point O ($J_2=J_3=0$) in the parameter space, 
where $E_{\rm sep}=-1.5J_1$, and the value of 
$T_{\rm E}$ is 0.802$J_1/K_{\rm B}$ which is identical to that of $T_{\rm c}^{\rm NN}$. 
This is due to the fact that at this point the resulting spin interactions are defined 
on a bipartite graph or lattice and EW based on thermal energy in this case 
detects only the bipartite entanglement. 
\section{Discussion}
\label{discussion}
The spin-$\frac{1}{2}$ AFM Heisenberg hexagon with 
all-to-all exchange couplings is considered to 
investigate the variety of entanglement properties.
Four different locations, R$_1$, R$_2$, L and P have been identified 
where the nature of ground states are different while 
QPT occurs over the line L including the point P. 
By exploiting its six-fold rotational symmetry three different kinds of 
CNs, CN$_{\rm NN}$, CN$_{\rm NNN}$ and CN$_{\rm FN}$ are introduced and those give 
totally different results. 
Both $T_{\rm c}^{\rm NN}$ and $T_{\rm c}^{\rm FN}$ decrease with 
the increase of $J_2$ (Fig. \ref{twoplot}) and ultimately vanish over the lines  
$J_2/J_1=1$ and $J_3/J_1=0.87J_2/J_1+0.14$, respectively. 
Those observations reveal the fact that the frustration 
opposes the bipartite entanglement in this system. 
Multipartite entanglements of this model have been studied where susceptibility, 
fidelity and internal energy serve as the EWs. Multipartite entanglements survive 
up to the temperature, $T_{\rm c}$ which is always higher than the $T_{\rm c}^{\rm NN}$. 
Thus, the bipartite entanglement diminishes due to thermal agitation
more quickly than the multipartite entanglement. Frustration leads to the 
higher values of $T_{\rm c}$, so it favors the multipartite entanglements. 
Fidelity measurement indicates that this model exhibits no six-spin entanglement 
even in zero temperature. 
Survival of multipartite entanglement at finite temperature 
has been studied in terms of internal energy as EW. Entanglement is found to 
persist at high temperatures in this system in the vicinity to 
the point M, 
where value of $\Fcal$ is the maximum. 
It appears that frustration is responsible for the robustness of 
quantum entanglement against the thermal effect around this point.
Existence of the multipartite entanglement at finite temperatures 
is found in this system where the bipartite entanglement vanishes at non-zero temperatures.  
EW in terms of susceptibility can detect the existence of both bipartite and multipartite 
entanglements collectively at finite temperatures. 
On the other hand, EW in terms of internal energy can 
detect bipartite and multipartite entanglements separately for the cases  
when the spin interactions are defined 
on bipartite and non-bipartite graphs or lattices, respectively. 
For this model, EW based on internal energy  
detects only the bipartite entanglement at the point O in the parameter space, and 
that measures only the multipartite entanglement at the other point P. 
Therefore, at the point O,  $T_{\rm c}^{\rm NN}=T_{\rm E}$. 
Similarly, $T_{\rm c}$ becomes equal to  $T_{\rm E}$ only at the point P, where 
only multipartite entanglement survives and measured separately by the EWs based on 
susceptibility and internal energy. It further appears that EW based on 
internal energy detects the collective existence of both bipartite 
and multipartite entanglements everywhere in the $J_2$-$J_3$ parameter space 
except the points O and P. Therefore, development of 
more effective EWs is necessary for precise detection of different types of 
entanglements separately. 
Engineering of entangled quantum state at room temperature is a new challenge. 
So, the frustrated AFM spin models could shed light in this direction. 

The inelastic neutron scattering study on Cu$_3$WO$_6$ 
reveals that spin-1/2 Cu$^{2+}$ ions 
are arranged on the vertices of 
hexagons in its crystalline state \cite{Hase}. 
Dynamic structure factor 
predicts the magnitudes of $J_1$, $J_2$ and $J_3$  
are 78.5K, 50.4K and 40.0K, respectively. 
As they satisfy the relation $J_1+J_3>2J_2$, this system belongs to 
the region R$_1$ having the RVB ground state, $\Psi_{42}$. 
Position of this compound in the $J_2$-$J_3$ parameter 
space is identified by the point C. 
Estimations of various quantities for 
Cu$_3$WO$_6$ yield following values. $\Fcal=0.51$, CN$_{\rm NN}=0.43$, 
$k_{\rm B}T^{\rm NN}_{\rm c}/J_1=0.50$, $F=0.44$, $k_{\rm B}T_{\rm c}/J_1=2.20$, 
$E_{\rm sep}/J_1=-0.92$ and  $k_{\rm B}T_{\rm E}/J_1=1.33$. 
Hence, this material is no more suitable to yield thermally stable 
multipartite entanglement. 
Therefore, in our opinion synthesis of new AFM compounds 
whose compositions as well as structures are very close to Cu$_3$WO$_6$ 
or other one  
such that $J_2 \geq J_1+J_3 /2$ becomes necessary for the production of  
thermally stable multipartite entanglement. 
\section{ACKNOWLEDGMENTS}
MD acknowledges the UGC fellowship, no. 524067 (2014), India.  
AKG acknowledges a BRNS-sanctioned 
research project, no. 37(3)/14/16/2015, India. 
\section{Author contribution statement}
MD did the analytical work and AKG did the numerical work.  
The manuscript was prepared jointly by both the authors. 
\appendix
\section{ENERGY EIGENVALUES AND EIGENSTATES}
\label{eigensystem}
In this section, we provide the analytic expressions of all eigenvectors and corresponding eigenvalues of the 
Hamiltonian, (Eq. \ref{ham}).
All energy eigenvalues with definite values of $S_{\rm T}$, $S^z_{\rm T}$, $\lambda_r$ and $p$ 
have been enlisted in the Tab. I. 
To express the eigenstates following notations have been used.
\bea
 && \bra{\psi^3_n}=T^{n-1}\bra{3}\left(n=1\right),\; \bra{3}=\bra{\uparrow\uparrow\uparrow\uparrow\uparrow\uparrow},\nonumber \\
 &&\bra{\psi^2_n}=T^{n-1}\bra{2}\left(n=1,2,3,4,5,6\right),\; \bra{2}=\bra{\uparrow\uparrow\uparrow\uparrow\uparrow\downarrow},\nonumber \\
 && \bra{\psi^1_n}_0=T^{n-1}\bra{1}_0\left(n=1,2,3,4,5,6\right),\; \bra{1}_0=\bra{\uparrow\uparrow\uparrow\downarrow\downarrow\uparrow},\nonumber \\
 && \bra{\psi^1_n}_1=T^{n-1}\bra{1}_1\left(n=1,2,3,4,5,6\right),\; \bra{1}_1=\bra{\downarrow\uparrow\uparrow\uparrow\downarrow\uparrow},\nonumber \\
 && \bra{\psi^1_n}_2=T^{n-1}\bra{1}_2\left(n=1,2,3\right), \;\bra{1}_2=\bra{\downarrow\uparrow\uparrow\downarrow\uparrow\uparrow},\nonumber \\
 & & \bra{\psi^0_n}_0=T^{n-1}\bra{0}_0\left(n=1,2,3,4,5,6\right),\; \bra{0}_0=\bra{\uparrow\uparrow\uparrow\downarrow\downarrow\downarrow},\nonumber \\
 & & \bra{\psi^0_n}_1=T^{n-1}\bra{0}_1\left(n=1,2,3,4,5,6\right),\; \bra{0}_1=\bra{\uparrow\uparrow\downarrow\downarrow\uparrow\downarrow},\nonumber \\
 & & \bra{\psi^0_n}_2=T^{n-1}\bra{0}_2\left(n=1,2,3,4,5,6\right),\; \bra{0}_2=\bra{\uparrow\downarrow\uparrow\downarrow\downarrow\uparrow},\nonumber \\
  & & \bra{\psi^0_n}_3=T^{n-1}\bra{0}_3\left(n=1,2\right), \;\bra{0}_3=\bra{\uparrow\downarrow\uparrow\downarrow\uparrow\downarrow},\nonumber \\
   & &  \bra{\psi^{-1}_n}_0=T^{n-1}\bra{-1}_0\left(n=1,2,3,4,5,6\right), \;\bra{-1}_0=\bra{\downarrow\downarrow\downarrow\uparrow\uparrow\downarrow},\nonumber \\
 && \bra{\psi^{-1}_n}_1=T^{n-1}\bra{-1}_1\left(n=1,2,3,4,5,6\right), \;\bra{-1}_1=\bra{\uparrow\downarrow\downarrow\downarrow\uparrow\downarrow},\nonumber \\
 && \bra{\psi^{-1}_n}_2=T^{n-1}\bra{-1}_2\left(n=1,2,3\right), \;\bra{-1}_2=\bra{\uparrow\downarrow\downarrow\uparrow\downarrow\downarrow},\nonumber \\
  && \bra{\psi^{-2}_n}=T^{n-1}\bra{-2}\left(n=1,2,3,4,5,6\right), \;\bra{-2}=\bra{\downarrow\downarrow\downarrow\downarrow\downarrow\uparrow},\nonumber \\
  && \bra{\psi^{-3}_n}=T^{n-1}\bra{-3}\left(n=1\right),\; \bra{-3}=\bra{\downarrow\downarrow\downarrow\downarrow\downarrow\downarrow}.\nonumber
\eea
 Here $T$ is a unitary cyclic right shift operator such that 
 $T\bra{abcdef}=\bra{fabcde} $, where 
$\bra{abcdef}=\bra{a}\otimes\bra{b}\otimes\bra{c}\otimes\bra{d}\otimes\bra{e}\otimes\bra{f}$.
  All the energy eigenstates are enlisted in the Tab. II.
 
\begin{table}[h]
\label{table_eigenvalues}
\caption{Energy eigenvalues of the 
spin-1/2  $J_1$-$J_2$-$J_3$ Heisenberg hexagon}
\def\arraystretch{0.78}
  \tabcolsep4pt\begin{tabular}{|c|c|c|c|c||c|c|c|c|c|}
\hline
 
$\boldsymbol{S_{\rm T}}$& $\boldsymbol{S^z_{\rm T}}$ &  \textbf{Energy eigenvalues}  & $\boldsymbol{\lambda_r}$ & $\boldsymbol{p}$ &$\boldsymbol{S_{\rm T}}$& $\boldsymbol{S^z_{\rm T}}$ &  \textbf{Energy eigenvalues}  & $\boldsymbol{\lambda_r}$ & $\boldsymbol{p}$    \\  
\hline
 3 & 3 & $E_1 =$  $\frac{3}{2}(J_1+J_2+\frac{1}{2}J_3)$ & 1 & 1 & 1 &  0    & $E_{33} = $  $ \frac{1}{2}(J_1-3J_2-\frac{1}{2}J_3)$ & -1 & 1 \\
\hline
3 & 2 & $ E_2 =$  $\frac{3}{2}(J_1+J_2+\frac{1}{2}J_3)$ & 1 & 1 & 1 &  0   & $E_{34} = $  $ -\frac{1}{4}(J_1+3J_2+J_3-d_2)$  & 1 & 3 \\
\hline
2 &  2   & $ E_3 = $  $ J_1-\frac{1}{4}J_3$ & -1 & 3 & 1 &  0   & $E_{35} = $  $ -\frac{1}{4}(J_1+3J_2+J_3+d_2)$  & 1 & 3 \\
\hline
2 &  2   & $E_4 = $  $ J_1-\frac{1}{4}J_3$ & -1 & 3 &  1 &  0  & $E_{36} = $  $ -\frac{1}{4}(J_1+3J_2+J_3-d_2)$  & 1 & 3 \\
\hline
2 &  2  & $E_5 =$  $ \frac{1}{2}(-J_1+3J_2-\frac{1}{2}J_3)$ & -1 & 1 & 1 &  0   & $E_{37} = $  $ -\frac{1}{4}(J_1+3J_2+J_3+d_2)$  & 1 & 3 \\
\hline
2 &  2    & $E_6 = $  $ \frac{3}{4}J_3$& 1 & 3 & 0 &  0   & $E_{38} = $  $ \frac{3}{2}(-J_1-J_2+\frac{1}{2}J_3)$  & 1 & 1 \\
\hline
2 &  2   & $E_7 = $  $ \frac{3}{4}J_3$& 1 & 3 & 0 &  0  & $E_{39} = $  $- \frac{1}{2}(J_1+3J_2+\frac{1}{2}J_3)$  & -1 & 3 \\   
\hline
3 &  1   & $E_8 = $  $ \frac{3}{2}(J_1+J_2+\frac{1}{2}J_3)$ & 1 & 1 & 0 &  0   & $E_{40} = $  $ -\frac{1}{2}(J_1+3J_2+\frac{1}{2}J_3)$  & -1 & 3 \\
\hline
2 &  1  & $E_9 = $  $ J_1-\frac{1}{4}J_3$& -1 & 3 &  0 &  0  & $E_{41} = $  $ -J_1-\frac{5}{4}J_3+\frac{1}{2}d_3$  & -1 & 1  \\
\hline
2 &  1    & $E_{10} = $  $ J_1-\frac{1}{4}J_3$ & -1 & 3 &  0 &  0  & $E_{42} = $  $ -J_1-\frac{5}{4}J_3-\frac{1}{2}d_3$  & -1 & 1 \\
\hline
2 &  1    & $E_{11} = $  $ \frac{3}{4}J_3$ & 1 & 3 &   3 & -1  & $E_{43} = $  $ \frac{3}{2}(J_1+J_2+\frac{1}{2}J_3)$ & 1 & 1  \\
\hline
2 & 1    & $E_{12} = $  $ \frac{3}{4}J_3$ & 1 & 3 & 2 &  -1    & $E_{44} = $  $ J_1-\frac{1}{4}J_3$ & -1 & 3\\
\hline
2 &  1   & $E_{13} = $  $ \frac{1}{2}(-J_1+3J_2-\frac{1}{2}J_3)$& -1 & 1 & 2 &  -1    & $E_{45} = $  $ J_1-\frac{1}{4}J_3$ & -1 & 3 \\
\hline
1 & 1   & $E_{14} = $  $ -J_1-\frac{1}{4}J_3+\frac{1}{2}d_1$ & 1 & 1 & 2 &  -1 & $E_{46} = $  $ \frac{3}{4}J_3$ & 1 & 3   \\
\hline
1 &  1  & $E_{15} = $  $ -J_1-\frac{1}{4}J_3-\frac{1}{2}d_1$ & 1 & 1 & 2 &  -1   &  $E_{47} = $  $ \frac{3}{4}J_3$ & 1 & 3  \\
\hline
1 &  1    & $E_{16} = $  $ -J_1-\frac{1}{4}J_3$ & -1 & 3 & 2 & -1   & $E_{48} = $  $ \frac{1}{2}(-J_1+3J_2-\frac{1}{2}J_3)$ & -1 & 1 \\
\hline
1 &  1    & $E_{17} = $  $ -J_1-\frac{1}{4}J_3$ & -1 & 3 & 1 &  -1   &  $E_{49} = $  $ -J_1-\frac{1}{4}J_3+\frac{1}{2}d_1$ & 1 & 1  \\
\hline
1 &  1    & $E_{18} = $  $ \frac{1}{2}(J_1-3J_2-\frac{1}{2}J_3)$ & -1 & 1 & 1 &  -1   & $E_{50} = $  $ -J_1-\frac{1}{4}J_3-\frac{1}{2}d_1$ & 1 & 1   \\
\hline
1 &  1  & $E_{19} = $  $ -\frac{1}{4}(J_1+3J_2+J_3-d_2)$ & 1 & 3 & 1 & -1  & $E_{51} = $  $ -J_1-\frac{1}{4}J_3$ & -1 & 3 \\
\hline
1 &  1   & $E_{20} = $  $- \frac{1}{4}(J_1+3J_2+J_3+d_2)$ & 1 & 3 & 1 &  -1    & $E_{52} = $  $ -J_1-\frac{1}{4}J_3$ & -1 & 3 \\
\hline
1 &  1   & $E_{21} = $  $- \frac{1}{4}(J_1+3J_2+J_3-d_2)$ & 1 & 3 & 1 &  -1    & $E_{53} = $  $ \frac{1}{2}(J_1-3J_2-\frac{1}{2}J_3)$  & -1 & 1  \\
\hline
1 &  1  & $E_{22} = $  $- \frac{1}{4}(J_1+3J_2+J_3+d_2)$ & 1 & 3 & 1 &  -1   & $E_{54} = $  $ -\frac{1}{4}(J_1+3J_2+J_3-d_2)$  & 1 & 3 \\
\hline
3 &  0   & $E_{23} = $  $ \frac{3}{2}(J_1+J_2+\frac{1}{2}J_3)$ & 1 & 1 &  1 &  -1   & $E_{55} = $  $ -\frac{1}{4}(J_1+3J_2+J_3+d_2)$  & 1 & 3 \\
\hline
2 &  0    & $E_{24} = $  $ J_1-\frac{1}{4}J_3$ & -1 & 3 & 1 & -1  & $E_{56} = $  $ -\frac{1}{4}(J_1+3J_2+J_3-d_2)$  & 1 & 3 \\ 
\hline
2 &  0    & $E_{25} = $  $ J_1-\frac{1}{4}J_3$ & -1 & 3 &  1 & -1  & $E_{57} = $  $ -\frac{1}{4}(J_1+3J_2+J_3+d_2)$  & 1 & 3\\
\hline
2 &  0    & $E_{26} = $  $ \frac{3}{4}J_3$ & 1 & 3 & 3 & -2 & $ E_{58} =$  $\frac{3}{2}(J_1+J_2+\frac{1}{2}J_3)$ & 1 & 1 \\
\hline
2 & 0   & $E_{27} = $  $ \frac{3}{4}J_3$ & 1 & 3 & 2 &  -2  & $ E_{59} = $  $ J_1-\frac{1}{4}J_3$  & -1 & 3 \\
\hline
2 &  0  & $E_{28} = $  $ \frac{1}{2}(-J_1+3J_2-\frac{1}{2}J_3)$ & -1 & 1  &  2 &  -2  & $E_{60} = $  $ J_1-\frac{1}{4}J_3$  & -1 & 3\\
\hline
1 &  0   & $E_{29} = $  $ -J_1-\frac{1}{4}J_3+\frac{1}{2}d_1$ & 1 & 1 & 2 & -2  & $E_{61} =$  $ \frac{1}{2}(-J_1+3J_2-\frac{1}{2}J_3)$  & -1 & 1\\
\hline
 1 &  0  & $E_{30} = $  $ -J_1-\frac{1}{4}J_3-\frac{1}{2}d_1$ & 1 & 1 & 2 & -2    & $E_{62} = $  $ \frac{3}{4}J_3$ & 1 & 3 \\
 \hline
 1 &  0  & $E_{31} = $  $ -J_1-\frac{1}{4}J_3$ & -1 & 3 & 2 &  -2   & $E_{63} = $  $ \frac{3}{4}J_3$ & 1 & 3 \\
 \hline
  1 &  0  & $E_{32} = $  $ -J_1-\frac{1}{4}J_3$ & -1 & 3 & 3 &  -3 & $E_{64} =$  $\frac{3}{2}(J_1+J_2+\frac{1}{2}J_3)$ & 1 & 1\\
  \hline
\end{tabular}

\end{table}

\begin{table}[h]
\def\arraystretch{0.5}
 \tabcolsep0.01pt\begin{tabular}{ll}
 $d_1=\sqrt{5J_1^2+9J_2^2+4J_3^2-10J_1J_2-8J_2J_3}$, &$d_2=\sqrt{17J_1^2+9J_2^2+16J_3^2-10J_1J_2-24J_1J_3-8J_2J_3}$, \\
\\
 $d_3=\sqrt{13J_1^2+9J_2^2+4J_3^2-18J_1J_2-8J_1J_3}$. \\ 
\\                                                    
\end{tabular}
\end{table} 


\begin{table}[h]
\label{table_eigenstates}
\caption{Eigenstates of the 
spin-1/2  $J_1$-$J_2$-$J_3$ Heisenberg hexagon}
 \tabcolsep4pt\begin{tabular}{|l|l|}

\hline

$\boldsymbol{S^z_{\rm T}}$& \textbf{Eigenstates}\\
\hline

3 & $\Psi_1=$    \bra{\psi^3_n} \\
\hline
2 &$\Psi_2=$  $\frac{1}{\sqrt{6}}$  $\sum\limits_{n=1}^{6}  \bra{\psi^2_n}$ \\  
\hline
2 & $\Psi_3=$  $\frac{1}{\sqrt{12}}$  $\left(\sum\limits_{n=1}^{6}\left(-1\right)^{n-1}  \bra{\psi^2_n}+3 \sum\limits_{n=3,6} \left(-1\right)^{n} \bra{\psi^2_n}\right)$   \\
\hline
2 &$\Psi_4=$  $ \frac{1}{2} $   $\left(\sum\limits_{n=1}^{2}  \bra{\psi^2_n}-\sum\limits_{n=4}^{5}  \bra{\psi^2_n}\right)$  \\
\hline
2 &$\Psi_5=$  $\frac{1}{\sqrt{6}}$ $\left(\sum\limits_{n=1}^{6}\left(-1\right)^{n-1}  \bra{\psi^2_n}\right)$  \\
\hline
2 &$\Psi_6=$  $\frac{1}{2}$ $\left(\sum\limits_{n=3,6}  \bra{\psi^2_n}-\sum\limits_{n=2,5}  \bra{\psi^2_n}\right)$   \\
\hline
2 &$\Psi_7=$  $\frac{1}{\sqrt{12}}$ $\left(3\sum\limits_{n=1,4}  \bra{\psi^2_n}- \sum\limits_{n=1}^{6} \bra{\psi^2_n}\right)$ \\
\hline
1 &$\Psi_8=$  $\frac{1}{\sqrt{15}}$  $\left(\sum\limits_{n=1}^{6} \left(  \bra{\psi^1_n}_0 +  \bra{\psi^1_n}_1\right) +  \sum\limits_{n=1}^{3}  \bra{\psi^1_n}_2 \right)$  \\
\hline
1 &$\Psi_9=$  $\frac{1}{4}$  $\left(\sum\limits_{n=1}^{6}\left(-1\right)^{n-1}  \bra{\psi^1_n}_0 +3 \sum\limits_{n=1,4} \left(-1\right)^{n} \bra{\psi^1_n}_0 +\sum\limits_{n=2}^{3}  \bra{\psi^1_n}_1 - \sum\limits_{n=5}^{6}  \bra{\psi^1_n}_1  \right)$    \\
\hline
1 &$\Psi_{10}=$  $\frac{1}{\sqrt{48}}$ $\left(3\sum\limits_{n=2}^{3}  \bra{\psi^1_n}_0 - 3\sum\limits_{n=5}^{6}  \bra{\psi^1_n}_0+\sum\limits_{n=1}^{6}\left(-1\right)^{n}  \bra{\psi^1_n}_1 +3 \sum\limits_{n=1,4} \left(-1\right)^{n-1} \bra{\psi^1_n}_1   \right)$   \\
\hline
1 &$\Psi_{11}=$  $\frac{1}{\sqrt{48}}$ $\left(3\sum\limits_{n=2,5}  \left( \bra{\psi^1_n}_0 +  \bra{\psi^1_n}_1\right)-\sum\limits_{n=1}^{6} \left( \bra{\psi^1_n}_0 +  \bra{\psi^1_n}_1\right) +2 \sum\limits_{n=2}^{3}  \bra{\psi^1_n}_2-4\bra{1}_2   \right)$    \\
\hline
1 &$\Psi_{12}=$  $\frac{1}{4}$ $\left(\sum\limits_{n=1,4} \left( \bra{\psi^1_n}_0 +  \bra{\psi^1_n}_1\right)-\sum\limits_{n=3,6} \left( \bra{\psi^1_n}_0 +  \bra{\psi^1_n}_1\right) +2 \sum\limits_{n=2}^{3}\left(-1\right)^{n}  \bra{\psi^1_n}_2  \right)$   \\
\hline
1 &$\Psi_{13}=$  $\frac{1}{\sqrt{6}}$ $\left(\sum\limits_{n=1}^{6}\left(-1\right)^{n-1}  \bra{\psi^1_n}_1 \right)$   \\
\hline
1 &$\Psi_{14}=$  $\frac{1}{\alpha_1\sqrt{6}}$ $\left(\sum\limits_{n=1}^{6}\left(C_{11}  \bra{\psi^1_n}_0+ \bra{\psi^1_n}_1\right)+\sqrt{2}C_{12}\sum\limits_{n=1}^{3} \bra{\psi^1_n}_2 \right)$ \\
\hline
1 &$\Psi_{15}=$  $\frac{1}{\alpha^\prime_1\sqrt{6}}$ $\left(\sum\limits_{n=1}^{6}\left(C^\prime_{11}  \bra{\psi^1_n}_0+ \bra{\psi^1_n}_1\right)+\sqrt{2}C^\prime_{12}\sum\limits_{n=1}^{3} \bra{\psi^1_n}_2 \right)$  \\
\hline
1 &$\Psi_{16}=$  $\frac{1}{\sqrt{48}}$ $\left(\sum\limits_{n=1}^{6}\left(-1\right)^{n}  \bra{\psi^1_n}_0 + 3\sum\limits_{n=1,4}\left(-1\right)^{n-1}  \bra{\psi^1_n}_0+ 3 \sum\limits_{n=2}^{3} \bra{\psi^1_n}_1 -3 \sum\limits_{n=5}^{6}  \bra{\psi^1_n}_1   \right)$  \\
\hline
1 &$\Psi_{17}=$  $\frac{1}{4}$ $\left(\sum\limits_{n=2}^{3} \bra{\psi^1_n}_0 -\sum\limits_{n=5}^{6}  \bra{\psi^1_n}_0+ \sum\limits_{n=1}^{6}\left(-1\right)^{n-1}  \bra{\psi^1_n}_1 + 3\sum\limits_{n=1,4}\left(-1\right)^{n}  \bra{\psi^1_n}_1  \right)$    \\
\hline
1 &$\Psi_{18}=$  $\frac{1}{\sqrt{6}}$ $\sum\limits_{n=1}^{6}\left(-1\right)^{n}  \bra{\psi^1_n}_0 $   \\
\hline
1 &$\Psi_{19}=$  $\frac{1}{\alpha_2\sqrt{12}}$ $\left(3\sum\limits_{n=2,5} \left(C_{21} \bra{\psi^1_n}_0+  \bra{\psi^1_n}_1\right)-\sum\limits_{n=1}^{6} \left(C_{21} \bra{\psi^1_n}_0+  \bra{\psi^1_n}_1\right)\right)$\\
 &$+\frac{1}{\alpha_2\sqrt{6}}C_{22}\left( 2\bra{1}_2-\sum\limits_{n=2}^{3} \bra{\psi^1_n}_2\right)$  \\
\hline

\end{tabular}

\end{table}

\begin{table}[h]

 \tabcolsep4pt\begin{tabular}{|l|l|}

\hline
$\boldsymbol{S^z_{\rm T}}$& \textbf{Eigenstates}\\
\hline
1 &$\Psi_{20}=$  $\frac{1}{\alpha^\prime_2\sqrt{12}}$ $\left(3\sum\limits_{n=2,5} \left(C^\prime_{21} \bra{\psi^1_n}_0+  \bra{\psi^1_n}_1\right)-\sum\limits_{n=1}^{6} \left(C^\prime_{21} \bra{\psi^1_n}_0+  \bra{\psi^1_n}_1\right)\right)$\\
 & $+\frac{1}{\alpha^\prime_2\sqrt{6}}C^\prime_{22}\left( 2\bra{1}_2-\sum\limits_{n=2}^{3} \bra{\psi^1_n}_2\right)$  \\
\hline
1 &$\Psi_{21}=$  $\frac{1}{2\alpha_2}$ $\left(\sum\limits_{n=1,4} \left(C_{21} \bra{\psi^1_n}_0+  \bra{\psi^1_n}_1\right)-\sum\limits_{n=3,6} \left(C_{21} \bra{\psi^1_n}_0+  \bra{\psi^1_n}_1\right)\right)$\\
 & $+\frac{1}{\alpha_2\sqrt{2}}C_{22}\sum\limits_{n=2}^{3}\left(-1\right)^{n-1} \bra{\psi^1_n}_2$  \\
\hline
1 &$\Psi_{22}=$  $\frac{1}{2\alpha^\prime_2}$ $\left(\sum\limits_{n=1,4} \left(C^\prime_{21} \bra{\psi^1_n}_0+  \bra{\psi^1_n}_1\right)-\sum\limits_{n=3,6} \left(C^\prime_{21} \bra{\psi^1_n}_0+  \bra{\psi^1_n}_1\right)\right)$\\
 & $\frac{1}{\alpha^\prime_2\sqrt{2}}C^\prime_{22}\sum\limits_{n=2}^{3}\left(-1\right)^{n-1} \bra{\psi^1_n}_2$  \\
\hline
0 &$\Psi_{23}=$  $\frac{1}{\sqrt{20}}$  $\left(\sum\limits_{n=1}^{6} \left(  \bra{\psi^0_n}_0 +  \bra{\psi^0_n}_1 + \bra{\psi^0_n}_2 \right) + \sum\limits_{n=1}^{2}  \bra{\psi^0_n}_3 \right)$ \\  
\hline
0 &$\Psi_{24}=$  $\frac{1}{\sqrt{24}}$ $\left(2 \sum\limits_{n=1}^{2}  \bra{\psi^0_n}_0-2\sum\limits_{n=4}^{5}  \bra{\psi^0_n}_0+\sum\limits_{n=2}^{3}  \left( \bra{\psi^0_n}_1 +  \bra{\psi^0_n}_2\right)-\sum\limits_{n=5}^{6} \left( \bra{\psi^0_n}_1 +  \bra{\psi^0_n}_2\right)   \right)$    \\
\hline
0 &$\Psi_{25}=$  $\frac{1}{\sqrt{72}}$ $\left(6\sum\limits_{n=3,6}\left(-1\right)^{n}  \bra{\psi^0_n}_0 +3\sum\limits_{n=1,4}\left(-1\right)^{n-1}  \left( \bra{\psi^0_n}_1 +  \bra{\psi^0_n}_2\right)\right)$\\
 & $+\frac{1}{\sqrt{72}}\sum\limits_{n=1}^{6}\left(-1\right)^{n} \left( \bra{\psi^0_n}_1 +  \bra{\psi^0_n}_2-2 \bra{\psi^0_n}_0\right)  $    \\
\hline
0 &$\Psi_{26}=$  $\frac{1}{\sqrt{8}}$ $\left(\sum\limits_{n=2,5}  \left( \bra{\psi^0_n}_2 -  \bra{\psi^0_n}_1\right)+\sum\limits_{n=3,6}\left( \bra{\psi^0_n}_1 -  \bra{\psi^0_n}_2\right) \right)$    \\
\hline
0 &$\Psi_{27}=$  $\frac{1}{\sqrt{24}}$ $\left(3\sum\limits_{n=1,4}  \left( \bra{\psi^0_n}_1 -  \bra{\psi^0_n}_2\right)+\sum\limits_{n=1}^{6}\left( \bra{\psi^0_n}_2 -  \bra{\psi^0_n}_1\right) \right)$    \\
\hline
0 &$\Psi_{28}=$  $\frac{1}{6}$ $\left(\sum\limits_{n=1}^{6}\left(-1\right)^{n-1}  \left( \bra{\psi^0_n}_0+ \bra{\psi^0_n}_1 +  \bra{\psi^0_n}_2\right)+3\sum\limits_{n=1}^{2}\left(-1\right)^{n-1} \bra{\psi^0_n}_3 \right)$    \\
\hline
0 &$\Psi_{29}=$  $\frac{1}{\alpha_3\sqrt{12}}$ $\left(\sum\limits_{n=1}^{6}\left(\sqrt{2}C_{31} \bra{\psi^0_n}_0+C_{32} \left( \bra{\psi^0_n}_1+ \bra{\psi^0_n}_2\right)\right)+\sqrt{6}\sum\limits_{n=1}^{2} \bra{\psi^0_n}_3 \right)$ \\
\hline
0 &$\Psi_{30}=$  $\frac{1}{\alpha^\prime_3\sqrt{12}}$ $\left(\sum\limits_{n=1}^{6}\left(\sqrt{2}C^\prime_{31} \bra{\psi^0_n}_0+C^\prime_{32} \left( \bra{\psi^0_n}_1+ \bra{\psi^0_n}_2\right)\right)+\sqrt{6}\sum\limits_{n=1}^{2} \bra{\psi^0_n}_3 \right)$ \\
\hline
0 &$\Psi_{31}=$  $\frac{1}{\sqrt{24}}$ $\left(\sum\limits_{n=1}^{6}\left(-1\right)^{n}  \left( \bra{\psi^0_n}_1- \bra{\psi^0_n}_2\right)+3\sum\limits_{n=1,4}\left(-1\right)^{n-1}\left( \bra{\psi^0_n}_1- \bra{\psi^0_n}_2\right) \right)$    \\
\hline
0 &$\Psi_{32}=$  $\frac{1}{\sqrt{8}}$ $\left(\sum\limits_{n=5}^{6}  \left( \bra{\psi^0_n}_2- \bra{\psi^0_n}_1\right)+\sum\limits_{n=2}^{3}\left( \bra{\psi^0_n}_1- \bra{\psi^0_n}_2\right) \right)$    \\
\hline
0 &$\Psi_{33}=$  $\frac{1}{\sqrt{12}}$ $\left(\sum\limits_{n=1}^{6}\left(-1\right)^{n-1} \left( \bra{\psi^0_n}_1 -  \bra{\psi^0_n}_2\right) \right)$    \\
\hline
0 &$\Psi_{34}=$  $\frac{C_4}{\alpha_4\sqrt{8}}$ $\left(\sum\limits_{n=3,6}\left( \bra{\psi^0_n}_1+ \bra{\psi^0_n}_2\right)-\sum\limits_{n=2,5}\left( \bra{\psi^0_n}_1+ \bra{\psi^0_n}_2\right)\right)$\\
& $+\frac{1}{2\alpha_4}\left(\sum\limits_{n=2,5}  \bra{\psi^0_n}_0-\sum\limits_{n=1,4} \bra{\psi^0_n}_0\right)$\\
\hline

\end{tabular}

\end{table}
\begin{table}[h]

 \tabcolsep4pt\begin{tabular}{|l|l|}
\hline
$\boldsymbol{S^z_{\rm T}}$& \textbf{Eigenstates}\\
\hline
0 &$\Psi_{35}=$  $\frac{C^\prime_4}{\alpha^\prime_4\sqrt{8}}$ $\left(\sum\limits_{n=3,6}\left( \bra{\psi^0_n}_1+ \bra{\psi^0_n}_2\right)-\sum\limits_{n=2,5}\left( \bra{\psi^0_n}_1+ \bra{\psi^0_n}_2\right)\right)$\\
& $+\frac{1}{2\alpha^\prime_4}\left(\sum\limits_{n=2,5}  \bra{\psi^0_n}_0-\sum\limits_{n=1,4} \bra{\psi^0_n}_0\right)$\\
\hline
0 &$\Psi_{36}=$  $\frac{C_4}{\alpha_4\sqrt{24}}$ $\left( 3 \sum\limits_{n=1,4} \left( \bra{\psi^0_n}_1+ \bra{\psi^0_n}_2\right)-\sum\limits_{n=1}^{6}\left( \bra{\psi^0_n}_1+ \bra{\psi^0_n}_2\right)\right)$\\
& $+\frac{1}{\alpha_4\sqrt{12}}\left(3\sum\limits_{n=3,6}\bra{\psi^0_n}_0-\sum\limits_{n=1}^{6} \bra{\psi^0_n}_0 \right)$\\
\hline
0 &$\Psi_{37}=$  $\frac{C^\prime_4}{\alpha^\prime_4\sqrt{24}}$ $\left( 3 \sum\limits_{n=1,4} \left( \bra{\psi^0_n}_1+ \bra{\psi^0_n}_2\right)-\sum\limits_{n=1}^{6}\left( \bra{\psi^0_n}_1+ \bra{\psi^0_n}_2\right)\right)$\\
& $+\frac{1}{\alpha^\prime_4\sqrt{12}}\left( 3\sum\limits_{n=3,6}  \bra{\psi^0_n}_0-\sum\limits_{n=1}^{6} \bra{\psi^0_n}_0 \right)$\\
\hline
0 &$\Psi_{38}=$  $\frac{1}{\sqrt{12}}$ $\left(\sum\limits_{n=1}^{6}  \left( \bra{\psi^0_n}_2- \bra{\psi^0_n}_1\right)\right)$ \\
\hline
0 &$\Psi_{39}=$  $\frac{1}{2}$ $\left(\sum\limits_{n=3,6} \left(-1\right)^{n}  \bra{\psi^0_n}_0+\sum\limits_{n=1,4}\left(-1\right)^{n} \left( \bra{\psi^0_n}_1+ \bra{\psi^0_n}_2\right)\right)$\\
 &$+\frac{1}{6}\sum\limits_{n=1}^{6}\left(-1\right)^{n-1}\left( \bra{\psi^0_n}_0+ \bra{\psi^0_n}_1+ \bra{\psi^0_n}_2\right)$\\
\hline
0 &$\Psi_{40}=$  $\frac{1}{\sqrt{12}}$ $\left(\sum\limits_{n=1}^{2}  \bra{\psi^0_n}_0-\sum\limits_{n=4}^{5}  \bra{\psi^0_n}_0+\sum\limits_{n=5}^{6}  \left( \bra{\psi^0_n}_1+ \bra{\psi^0_n}_2\right)-\sum\limits_{n=2}^{3}  \left( \bra{\psi^0_n}_1+ \bra{\psi^0_n}_2\right)\right)$ \\
\hline
0 &$\Psi_{41}=$  $\frac{1}{\alpha_5\sqrt{12}}$ $\left(\sum\limits_{n=1}^{6} \left(-1\right)^{n-1} \left(\sqrt{2}C_{51} \bra{\psi^0_n}_0+ \bra{\psi^0_n}_1+ \bra{\psi^0_n}_2\right)+\sqrt{6}C_{52}\sum\limits_{n=1}^{2} \left(-1\right)^{n-1}  \bra{\psi^0_n}_3\right)$\\
\hline
0 &$\Psi_{42}=$  $\frac{1}{\alpha^\prime_5\sqrt{12}}$ $\left(\sum\limits_{n=1}^{6} \left(-1\right)^{n-1} \left(\sqrt{2}C^\prime_{51} \bra{\psi^0_n}_0+ \bra{\psi^0_n}_1+ \bra{\psi^0_n}_2\right)+\sqrt{6}C^\prime_{52}\sum\limits_{n=1}^{2} \left(-1\right)^{n-1}  \bra{\psi^0_n}_3\right)$\\
\hline
-1 &$\Psi_{43}=$  $\frac{1}{\sqrt{15}}$  $\left(\sum\limits_{n=1}^{6} \left(  \bra{\psi^{-1}_n}_0 +  \bra{\psi^{-1}_n}_1\right) +  \sum\limits_{n=1}^{3}  \bra{\psi^{-1}_n}_2 \right)$  \\
\hline
-1 &$\Psi_{44}=$  $\frac{1}{4}$  $\left(\sum\limits_{n=1}^{6}\left(-1\right)^{n-1}  \bra{\psi^{-1}_n}_0 +3 \sum\limits_{n=1,4} \left(-1\right)^{n} \bra{\psi^{-1}_n}_0 +\sum\limits_{n=2}^{3}  \bra{\psi^{-1}_n}_1 - \sum\limits_{n=5}^{6}  \bra{\psi^{-1}_n}_1  \right)$    \\
\hline
-1 &$\Psi_{45}=$  $\frac{1}{\sqrt{48}}$ $\left(3\sum\limits_{n=2}^{3}  \bra{\psi^{-1}_n}_0 - 3\sum\limits_{n=5}^{6}  \bra{\psi^{-1}_n}_0+\sum\limits_{n=1}^{6}\left(-1\right)^{n}  \bra{\psi^{-1}_n}_1 +3 \sum\limits_{n=1,4} \left(-1\right)^{n-1} \bra{\psi^{-1}_n}_1   \right)$   \\
\hline
-1 &$\Psi_{46}=$  $\frac{1}{\sqrt{48}}$ $\left(3\sum\limits_{n=2,5}  \left( \bra{\psi^{-1}_n}_0 +  \bra{\psi^{-1}_n}_1\right)-\sum\limits_{n=1}^{6} \left( \bra{\psi^{-1}_n}_0 +  \bra{\psi^{-1}_n}_1\right) +2 \sum\limits_{n=2}^{3}  \bra{\psi^{-1}_n}_2-4\bra{-1}_2   \right)$    \\
\hline
-1 &$\Psi_{47}=$  $\frac{1}{4}$ $\left(\sum\limits_{n=1,4}  \left( \bra{\psi^{-1}_n}_0 +  \bra{\psi^{-1}_n}_1\right)-\sum\limits_{n=3,6} \left( \bra{\psi^{-1}_n}_0 +  \bra{\psi^{-1}_n}_1\right) +2 \sum\limits_{n=2}^{3}\left(-1\right)^{n}  \bra{\psi^{-1}_n}_2  \right)$   \\
\hline
-1 &$\Psi_{48}=$  $\frac{1}{\sqrt{6}}$ $\left(\sum\limits_{n=1}^{6}\left(-1\right)^{n-1}  \bra{\psi^{-1}_n}_1 \right)$   \\
\hline
-1 &$\Psi_{49}=$  $\frac{1}{\alpha_1\sqrt{6}}$ $\left(\sum\limits_{n=1}^{6}\left(C_{11}  \bra{\psi^{-1}_n}_0+ \bra{\psi^{-1}_n}_1\right)+\sqrt{2}C_{12}\sum\limits_{n=1}^{3} \bra{\psi^{-1}_n}_2 \right)$ \\
\hline
-1 &$\Psi_{50}=$  $\frac{1}{\alpha^\prime_1\sqrt{6}}$ $\left(\sum\limits_{n=1}^{6}\left(C^\prime_{11}  \bra{\psi^{-1}_n}_0+ \bra{\psi^{-1}_n}_1\right)+\sqrt{2}C^\prime_{12}\sum\limits_{n=1}^{3} \bra{\psi^{-1}_n}_2 \right)$  \\
\hline

\end{tabular}

\end{table}           
\begin{table}[h]
 \tabcolsep4pt\begin{tabular}{|l|l|}
\hline
$\boldsymbol{S^z_{\rm T}}$& \textbf{Eigenstates}\\
\hline
-1 &$\Psi_{51}=$  $\frac{1}{\sqrt{48}}$ $\left(\sum\limits_{n=1}^{6}\left(-1\right)^{n}  \bra{\psi^{-1}_n}_0 + 3\sum\limits_{n=1,4}\left(-1\right)^{n-1}  \bra{\psi^{-1}_n}_0+ 3 \sum\limits_{n=2}^{3} \bra{\psi^{-1}_n}_1 -3 \sum\limits_{n=5}^{6}  \bra{\psi^{-1}_n}_1   \right)$  \\
\hline
-1 &$\Psi_{52}=$  $\frac{1}{4}$ $\left(\sum\limits_{n=2}^{3} \bra{\psi^{-1}_n}_0 -\sum\limits_{n=5}^{6}  \bra{\psi^{-1}_n}_0+ \sum\limits_{n=1}^{6}\left(-1\right)^{n-1}  \bra{\psi^{-1}_n}_1 + 3\sum\limits_{n=1,4}\left(-1\right)^{n}  \bra{\psi^{-1}_n}_1  \right)$    \\
\hline
-1 &$\Psi_{53}=$  $\frac{1}{\sqrt{6}}$ $\left(\sum\limits_{n=1}^{6}\left(-1\right)^{n}  \bra{\psi^{-1}_n}_0 \right)$   \\
\hline
-1 &$\Psi_{54}=$  $\frac{1}{\alpha_2\sqrt{12}}$ $\left(3\sum\limits_{n=2,5} \left(C_{21} \bra{\psi^{-1}_n}_0+  \bra{\psi^{-1}_n}_1\right)-\sum\limits_{n=1}^{6} \left(C_{21} \bra{\psi^{-1}_n}_0+  \bra{\psi^{-1}_n}_1\right)\right)$  \\
 & $+\frac{1}{\alpha_2\sqrt{6}}C_{22}\left( 2\bra{-1}_2-\sum\limits_{n=2}^{3} \bra{\psi^{-1}_n}_2\right)$\\
\hline
-1 &$\Psi_{55}=$ $\frac{1}{\alpha^\prime_2\sqrt{12}}$ $\left(3\sum\limits_{n=2,5} \left(C^\prime_{21} \bra{\psi^{-1}_n}_0+  \bra{\psi^{-1}_n}_1\right)-\sum\limits_{n=1}^{6} \left(C^\prime_{21} \bra{\psi^{-1}_n}_0+  \bra{\psi^{-1}_n}_1\right)\right)$  \\
 & $+\frac{1}{\alpha^\prime_2\sqrt{6}}C^\prime_{22}\left( 2\bra{-1}_2-\sum\limits_{n=2}^{3} \bra{\psi^{-1}_n}_2\right)$\\
\hline
-1 &$\Psi_{56}=$  $\frac{1}{2\alpha_2}$ $\left(\sum\limits_{n=1,4} \left(C_{21} \bra{\psi^{-1}_n}_0+  \bra{\psi^{-1}_n}_1\right)-\sum\limits_{n=3,6} \left(C_{21} \bra{\psi^{-1}_n}_0+  \bra{\psi^{-1}_n}_1\right)\right)$  \\
 & $+\frac{1}{\alpha_2\sqrt{2}}C_{22}\sum\limits_{n=2}^{3}\left(-1\right)^{n-1} \bra{\psi^{-1}_n}_2 $\\ 
\hline
-1 &$\Psi_{57}=$ $\frac{1}{2\alpha^\prime_2}$ $\left(\sum\limits_{n=1,4} \left(C^\prime_{21} \bra{\psi^{-1}_n}_0+  \bra{\psi^{-1}_n}_1\right)-\sum\limits_{n=3,6} \left(C^\prime_{21} \bra{\psi^{-1}_n}_0+  \bra{\psi^{-1}_n}_1\right)\right)$  \\
 & $+\frac{1}{\alpha^\prime_2\sqrt{2}}C^\prime_{22}\sum\limits_{n=2}^{3}\left(-1\right)^{n-1} \bra{\psi^{-1}_n}_2 $\\
\hline
-2 & $\Psi_{58}=$  $\frac{1}{\sqrt{6}}$  $\sum\limits_{n=1}^{6}  \bra{\psi^{-2}_n} $ \\  
\hline
-2 & $\Psi_{59}=$  $\frac{1}{\sqrt{12}}$  $\left(\sum\limits_{n=1}^{6}\left(-1\right)^{n-1}  \bra{\psi^{-2}_n} +3 \sum\limits_{n=3,6} \left(-1\right)^{n} \bra{\psi^{-2}_n} \right)$   \\
\hline
-2 &$\Psi_{60}=$  $ \frac{1}{2} $   $\left(\sum\limits_{n=1}^{2}  \bra{\psi^{-2}_n} -\sum\limits_{n=4}^{5}  \bra{\psi^{-2}_n} \right)$  \\   
\hline
-2 &$\Psi_{61}=$  $\frac{1}{\sqrt{6}}$ $\sum\limits_{n=1}^{6}\left(-1\right)^{n-1}  \bra{\psi^{-2}_n} $  \\
\hline
-2 &$\Psi_{62}=$  $\frac{1}{2}$ $\left(\sum\limits_{n=3,6}  \bra{\psi^{-2}_n} -\sum\limits_{n=2,5}  \bra{\psi^{-2}_n} \right)$   \\
\hline
-2 &$\Psi_{63}=$  $\frac{1}{\sqrt{12}}$ $\left(3\sum\limits_{n=1,4}  \bra{\psi^{-2}_n} - \sum\limits_{n=1}^{6}  \bra{\psi^{-2}_n} \right)$ \\
\hline
-3 &$\Psi_{64}=$    \bra{\psi^{-3}_n} \\
\hline
\end{tabular}
\end{table}   


\begin{table}[h]
\def\arraystretch{0.5}
  \tabcolsep1 pt\begin{tabular}{ll}
     $C_{11}=$  $\frac{\left(2J_1+2J_3\right)\left(J_1-J_2+2J_3-d_1\right)-8J_1J_2}{8J_2^2-\left(3J_1+J_2-d_1\right)\left(J_1-J_2+2J_3-d_1\right)}$, &
$C_{12}=$  $\frac{2\sqrt{2}\left(J_2\left(2J_1+2J_3\right)-J_1\left(3J_1+J_2-d_1\right)\right)}{\left(3J_1+J_2-d_1\right)\left(J_1-J_2+2J_3-d_1\right)-8J_2^2}$,\\ 
\\
$C^\prime_{11}=$  $\frac{\left(2J_1+2J_3\right)\left(J_1-J_2+2J_3+d_1\right)-8J_1J_2}{8J_2^2-\left(3J_1+J_2+d_1\right)\left(J_1-J_2+2J_3+d_1\right)}$, &
$C^\prime_{12}=$  $\frac{2\sqrt{2}\left(J_2\left(2J_1+2J_3\right)-J_1\left(3J_1+J_2+d_1\right)\right)}{\left(3J_1+J_2+d_1\right)\left(J_1-J_2+2J_3+d_1\right)-8J_2^2}$,\\
\\
$C_{21}=$  $\frac{8J_1J_2-2\left(2J_3-J_1\right)\left(-J_1+J_2+4J_3-d_2\right)}{\left(3J_1-J_2-d_2\right)\left(-J_1+J_2+4J_3-d_2\right)-8J_2^2}$, &
$C_{22}=$  $\frac{2\sqrt{2}\left(J_1\left(3J_1-J_2-d_2\right)-2J_2\left(2J_3-J_1\right)\right)}{\left(3J_1-J_2-d_2\right)\left(-J_1+J_2+4J_3-d_2\right)-8J_2^2}$,\\ 
\\
$C^\prime_{21}=$  $\frac{8J_1J_2-2\left(2J_3-J_1\right)\left(-J_1+J_2+4J_3+d_2\right)}{\left(3J_1-J_2+d_2\right)\left(-J_1+J_2+4J_3+d_2\right)-8J_2^2}$, &
$C^\prime_{22}=$  $\frac{2\sqrt{2}\left(J_1\left(3J_1-J_2+d_2\right)-2J_2\left(2J_3-J_1\right)\right)}{\left(3J_1-J_2+d_2\right)\left(-J_1+J_2+4J_3+d_2\right)-8J_2^2}$,\\ 
\\   
$C_{31}=$  $\frac{\sqrt{3}\left(J_3\left(3J_1+J_2+2J_3-d_1\right)-2J_1\left(2J_2+J_1\right)\right)}{2\left(2J_2+J_1\right)^2-\left(3J_1+J_2+2J_3-d_1\right)\left(3J_1-J_2+J_3-d_1\right) }$, &
$C_{32}=$  $\frac{\sqrt{6}\left(J_1\left(3J_1-J_2+J_3-d_1\right)-J_3\left(2J_2+J_1\right)\right)}{2\left(2J_2+J_1\right)^2-\left(3J_1+J_2+2J_3-d_1\right)\left(3J_1-J_2+J_3-d_1\right)}$, \\
\\
$C^\prime_{31}=$  $\frac{\sqrt{3}\left(J_3\left(3J_1+J_2+2J_3+d_1\right)-2J_1\left(2J_2+J_1\right)\right)}{2\left(2J_2+J_1\right)^2-\left(3J_1+J_2+2J_3+d_1\right)\left(3J_1-J_2+J_3+d_1\right)}$, &
$C^\prime_{32}=$  $\frac{\sqrt{6}\left(J_1\left(3J_1-J_2+J_3+d_1\right)-J_3\left(2J_2+J_1\right)\right)}{2\left(2J_2+J_1\right)^2-\left(3J_1+J_2+2J_3+d_1\right)\left(3J_1-J_2+J_3+d_1\right)}$, \\
\\
$C_4=$  $\frac{2\sqrt{2}\left(J_1-J_2\right)}{3J_1+J_2-4J_3+d_2}$, &
$C^\prime_4=$  $\frac{2\sqrt{2}\left(J_1-J_2\right)}{3J_1+J_2-4J_3-d_2}$, \\
\\
$C_{51}=$  $\frac{\sqrt{2}\left(3J_1J_3-\left(2J_2-J_1\right)\left(-J_1+3J_2+J_3-d_3\right)\right)}{\left(3J_1-J_2-J_3-d_3\right)\left(-J_1+3J_2+J_3-d_3\right)-3J_3^2}$, &
$C_{52}=$  $\frac{\sqrt{6}\left(J_1\left(3J_1-J_2-J_3-d_3\right)-J_3\left(2J_2-J_1\right)\right)}{3J_3^2-\left(3J_1-J_2-J_3-d_3\right)\left(-J_1+3J_2+J_3-d_3\right)}$,\\ 
\\ 
$C^\prime_{51}=$  $\frac{\sqrt{2}\left(3J_1J_3-\left(2J_2-J_1\right)\left(-J_1+3J_2+J_3+d_3\right)\right)}{\left(3J_1-J_2-J_3+d_3\right)\left(-J_1+3J_2+J_3+d_3\right)-3J_3^2}$, &
$C^\prime_{52}=$  $\frac{\sqrt{6}\left(J_1\left(3J_1-J_2-J_3+d_3\right)-J_3\left(2J_2-J_1\right)\right)}{3J_3^2-\left(3J_1-J_2-J_3+d_3\right)\left(-J_1+3J_2+J_3+d_3\right)}$,\\ 
\\ 
$\alpha_1=$  $\sqrt{1+C_{11}^2+C_{12}^2}$, & $\alpha^\prime_1=$  $\sqrt{1+C^{\prime2}_{11}+C^{\prime2}_{12}}$, \\
\\
$\alpha_2=$  $\sqrt{1+C_{21}^2+C_{22}^2}$, & $\alpha^\prime_2=$  $\sqrt{1+C^{\prime2}_{21}+C^{\prime2}_{22}}$,\\
\\
$\alpha_3=$  $\sqrt{1+C_{31}^2+C_{32}^2}$, & $\alpha^\prime_3=$  $\sqrt{1+C^{\prime2}_{31}+C^{\prime2}_{32}}$, \\
\\
$\alpha_4=$  $\sqrt{1+C_{4}^2}$, & $\alpha^\prime_4=$  $\sqrt{1+C^{\prime2}_{4}}$,\\
\\
$\alpha_5=$  $\sqrt{1+C_{51}^2+C_{52}^2}$, & $\alpha^\prime_5=$  $\sqrt{1+C^{\prime2}_{51}+C^{\prime2}_{52}}$.\\
\end{tabular}
\end{table} 


\clearpage

\begin{thebibliography}{99}
\bibitem{Amico}L. Amico, R. Fazio, A. Osterloh and V. Vedral, Rev. Mod. Phys. {\bf 80}, 517 (2008).
\bibitem{HorodeckisRMP}R. Horodecki, P. Horodecki, M. Horodecki and K. Horodecki, 
Rev. Mod. Phys. {\bf 81}, 865 (2009). 
\bibitem{Guuhne}O. G\"uuhne and G. Toth, Phys. Rep. {\bf 474}, 1 (2009).
\bibitem{Ekert}A. K. Ekert, Phys. Rev. Lett. {\bf 67}, 661 (1991).
\bibitem{Bennett1}C. H. Bennett and S. J. Wiesner,  Phys. Rev. Lett. {\bf 69}, 2881 (1992).
\bibitem{Bennett2}C. H. Bennett {\em et. al.}, Phys. Rev. Lett. {\bf 70}, 1895 (1993). 
\bibitem{QSTNE} G. M. Nikolopoulos and I. Jex, 
{\it Quantum State Transfer and Network Engineering}, 
Springer, Heidelberg (2014). 
\bibitem{Sachdev} S. Sachdev, {\it Quantum Phase Transitions}, 
Cambridge University Press, Cambridge,England, (1999). 
\bibitem{Wu} L. -A. Wu, M. S. Sarandy and D. A. Lidar, Phys. Rev. Lett. {\bf 93}, 250404 (2004). 
\bibitem{Somma} R. Somma, G. Ortiz, H. Barnum, E. Knill, and L. Viola, Phys.
Rev. A {\bf 70}, 042311 (2004). 
\bibitem{JVidal} J. Vidal, R. Mosseri, and J. Dukelsky, Phys. Rev. A {\bf 69},
054101 (2004). 
\bibitem{Procissi} D. Procissi {\em et. al.}, Phys. Rev. B. {\bf 69}, 094436 (2004). 
\bibitem{Bose}I. Bose and  A. Trivedi, Phys. Rev. A {\bf 72}, 022314 (2005).
\bibitem{Ghosh}S. Ghosh, T. F. Rosenbaum, G. Aeppli and S. N. Coppersmith, 
Nature (London) {\bf 425}, 48 (2003). 
\bibitem{Dowling}M. R. Dowling, A. C. Doherty, and S. D. Bartlett Phys. Rev. A {\bf 70}, 062113 (2004). 
\bibitem{Wiesniak} M. Wie\'{s}niak. V. Vedral. and \v{C}. Brunkner, 
New J. Phys. {\bf 7}, 258 (2005).
\bibitem{Krammer} P. Krammer {\em et. al.}, Phys. Rev. Lett. {\bf 103}, 100502 (2009).
\bibitem{Brukner}C. Brukner, V. Vedral and A Zeilinger, 
 Phys. Rev. A {\bf 73}, 012110 (2006). 
\bibitem{Korepin}B. -Q. Jin and V. E. Korepin, J. Stat. Phys.  {\bf 116}, 79 (2004).
\bibitem{Mezzadri}J. P. Keating and F. Mezzadri, Phys. Rev. Lett. {\bf 94}, 050501 (2005).
\bibitem{Damski}B. Damski and M. M. Rams, J. Phys. A: Math. Theor. {\bf 47}, 025303
(2014). 
\bibitem{Osborne} T. J. Osborne, and M. A. Nielsen, Phys. Rev. A {\bf 66}, 032110 (2002).
\bibitem{GVidal} G. Vidal, J. I. Latorre, E. Rico, and A. Kitaev, Phys. Rev. Lett.
{\bf 90}, 227902 (2003). 
\bibitem{Wooters_prl80} W. K. Wootters, Phys. Rev. Lett. {\bf 80}, 2245 (1998).
\bibitem{Wooters_pra63} K. M. O’Connor and W. K. Wootters, Phys. Rev. A {\bf 63}, 052302 (2001). 
\bibitem{Bruss} D. Bruss, N. Dutta, A. Ekert, L. C. Kwek, and C. Macchiavello, Phys. Rev. A {\bf 72}, 014301 (2005). 
\bibitem{Glaser} U. Glaser, H. Büttner, and H. Fehske, Phys. Rev. A {\bf 68}, 032318 (2003).
\bibitem{Osterloh} A. Osterloh, L. Amico, G. Falci, and R. Fazio, Nature (London) {\bf 416}, 608 (2002).
\bibitem{Souza} A. M. Souza {\em et. al.}, Phys. Rev. B. {\bf 79}, 054408 (2009). 
\bibitem{Sen} A. Sen (De) {\em et. al.}, Phys. Rev. Lett. {\bf 101}, 187202 (2008). 
\bibitem{Horodeckis}M. Horodecki, P. Horodecki and R. Horodecki, 
Phys. Lett. {\bf A 223}, 1 (1996).   
\bibitem{Terhal}B. M. Terhal, Phys. Lett. {\bf A 271}, 319 (2000). 
\bibitem{Wang} X. Wang, Phys. Rev. A {\bf 66}, 044305 (2002).
\bibitem{Sackett} C. A. Sackett {\em et. al.}, Nature (London) {\bf 404}, 256 (2000).
\bibitem{Bennett} C. H. Bennett {\em et. al.}, Phys. Rev. Lett. {\bf 76}, 722 (1996). 
\bibitem{Nelder_Mead} J. A. Nelder and R. Mead, Comput. J. {\bf 7}, 308 (1965). 
\bibitem{Hase}M. Hase {\em et. al.}, J. Phys. Soc. Jpn. {\bf 65}, 
372-375 (1996). 
\end{thebibliography}
\end{document}